\documentclass[aps,pra,reprint,amsmath,amssymb,superscriptaddress,longbibliography]{revtex4-1}
\usepackage[utf8]{inputenc}
\usepackage{graphicx}% Include figure files
\usepackage{dcolumn}% Align table columns on decimal point
\usepackage{bm}% bold math
\usepackage{amsmath}
\DeclareMathOperator{\arctanh}{arctanh}
\usepackage{color}
\usepackage{xspace}
\usepackage{ulem}
\usepackage{xfrac}
\usepackage{color,soul}
\usepackage{graphicx}
\usepackage[colorlinks=true,urlcolor=blue,citecolor=blue,linkcolor=blue,bookmarks=false,pdfstartview={FitH}]{hyperref}

\begin{document}

\title{\begin{small}This is a preprint of an article published in npj Quantum Materials. The final authenticated version is available online at: \href{url}{https://doi.org/10.1038/s41535-021-00351-4}\end{small}
	\vspace{2mm}
	\\
	Critical charge fluctuations and emergent coherence in a strongly correlated excitonic insulator}

\author{Pavel A. Volkov}
\email{pv184@physics.rutgers.edu}
\author{Mai Ye}
\email{mye@physics.rutgers.edu}
\affiliation{Department of Physics and Astronomy, Rutgers University, Piscataway, NJ 08854, USA}
\author{Himanshu Lohani}
\affiliation{Department of Physics, Technion - Israel Institute of Technology, Haifa 32000, Israel}
\author{Irena Feldman}
\affiliation{Department of Physics, Technion - Israel Institute of Technology, Haifa 32000, Israel}
\author{Amit Kanigel}
\affiliation{Department of Physics, Technion - Israel Institute of Technology, Haifa 32000, Israel}
\author{Girsh Blumberg}
\email{girsh@physics.rutgers.edu}
\affiliation{Department of Physics and Astronomy, Rutgers University, Piscataway, NJ 08854, USA}
\affiliation{National Institute of Chemical Physics and Biophysics, 12618 Tallinn, Estonia}

\date{\today}

\begin{abstract}
Excitonic insulator is a coherent electronic phase that results from the formation of a macroscopic population of bound particle-hole pairs – excitons. With only a few candidate materials known, the collective excitonic behavior is challenging to observe, being obscured by crystalline lattice effects. Here we use polarization-resolved Raman spectroscopy to reveal the quadrupolar excitonic mode in the candidate zero-gap semiconductor Ta$_2$NiSe$_5$ disentangling it from the lattice phonons. The excitonic mode pronouncedly softens close to the phase transition, showing its electronic character, while its coupling to non-critical lattice modes is shown to enhance the transition temperature. On cooling, we observe the gradual emergence of coherent superpositions of band states at the correlated insulator gap edge, with strong departures from mean-field theory predictions. Our results demonstrate the realization of a strongly correlated excitonic state in an equilibrium bulk material.
\end{abstract}

\maketitle
\noindent\section*{Introduction}

Attractive interactions between fermions are known to lead to a proliferation of bound pairs of particles at low temperatures causing a transition into superconducting or superfluid phases. In a semiconductor or a compensated semimetal, the Coulomb attraction between electrons and holes may induce a similar transition where electron-hole pairs, the excitons, develop macroscopic coherence~\cite{Keldysh1965,Ex1967,Ex1968,Ex1970}. The resulting state, characterized by an interaction-induced gap, has been dubbed as the excitonic insulator~\cite{jerome1967}. However, so far only a few materials have been identified as excitonic insulator candidates~\cite{neuensch1990,rossnagel2011,du2017,TS2017,li2019Sb,zhu2019graph} - possibly because its formation requires strong attraction or matching energy dispersions of the electron- and hole-like carriers~\cite{Keldysh1965}. These restrictions can be overcome by creating a non-equilibrium exciton population and cooling below their degeneracy temperature, in which case the coherent state may be observable only as a transient due to the finite lifetime of the excited state~\cite{snoke2002,deng2010,byrnes2014}. An equilibrium excitonic phase in a bulk material, on the other hand, would allow for a far wider range of questions to be asked and answered regarding the excitonic states of matter and their formation. 

\begin{figure*}
	\centering
	\includegraphics[width=14.5cm]{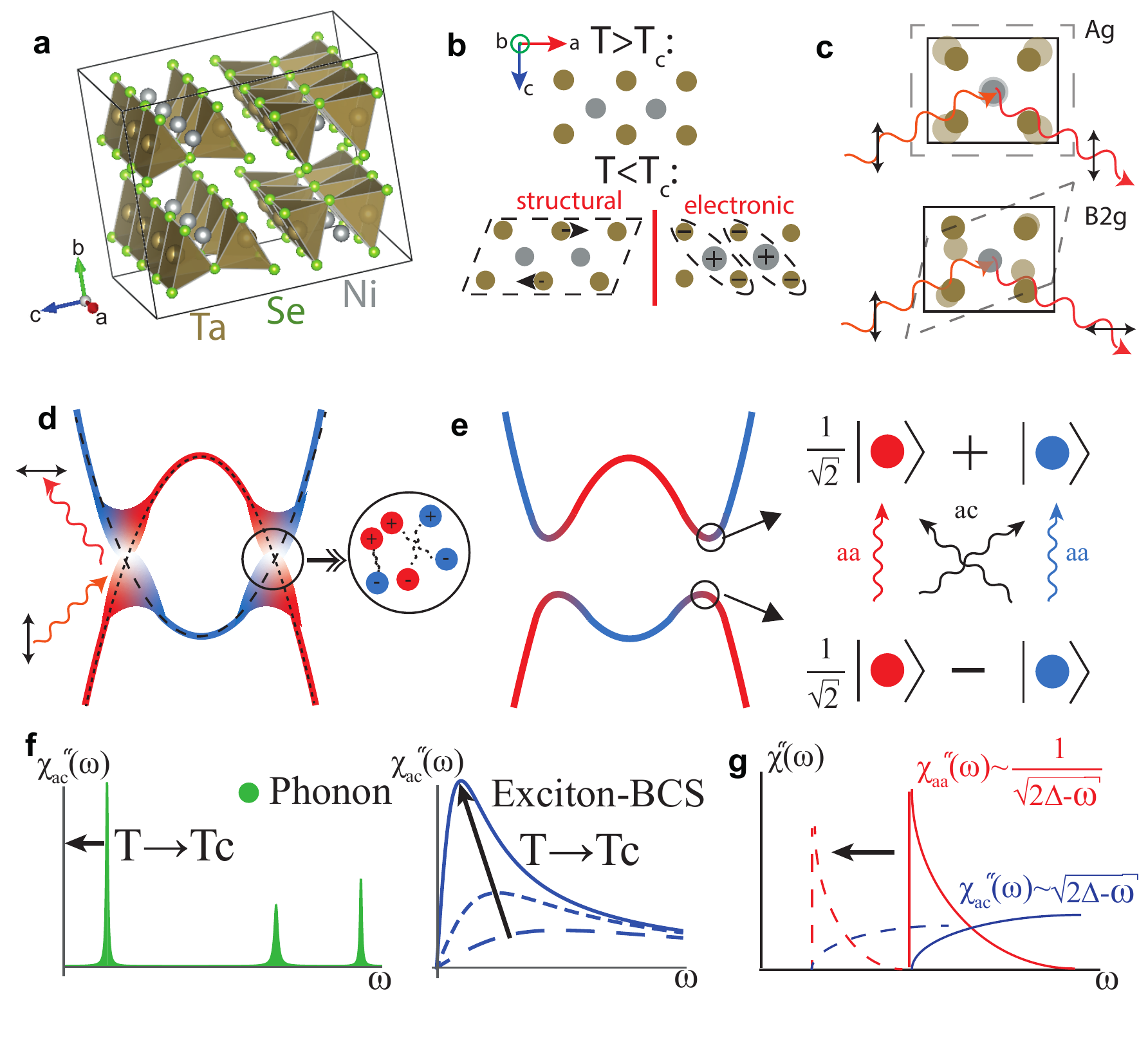}
	\caption{\textbf{Schematics of the excitonic insulator physics for Ta$_2$NiSe$_5$.} \textbf{a}, Crystal structure in the high-temperature orthorhombic phase. Three unit cells in $a$ direction are shown. \textbf{b}, Possible mechanisms of the phase transition: below $T_{\mathrm{c}}$, the symmetry of the lattice can be reduced due to structural distortion or coherent excitonic condensation. \textbf{c}, Illustration of the polarization-resolved Raman process: the $aa$ scattering geometry (top) probes excitations with the full symmetry of the lattice ($A_{\mathrm{g}}$), while the $ac$ scattering geometry (bottom) couples to the symmetry-breaking excitations with the symmetry of the $ac$-like quadrupole order parameter ($B_{\mathrm{2g}}$). \textbf{d} and \textbf{e}, Excitonic transition in a semimetal: above $T_{\mathrm{c}}$ \textbf{d} conduction (blue) and valence (red) bands cross at the Fermi level; pre-formed excitons are coupled to interband transitions forming an overdamped collective mode in the $B_{\mathrm{2g}}$ channel. 
		\textbf{e}, At low temperatures excitonic order hybridizes the bands, opening a spectral gap. At the gap edge, the eigenstates are equal-weight superpositions of the two bands, shown on the right. As a result, distinct interference effects occur for $ac$ (black arrows) and $aa$ (red/blue arrows) geometries. \textbf{f}, Expected Raman spectra for the $ac$ scattering geometry near $T_{\mathrm{c}}$. Left: for a structural transition one of the phonon modes softens to zero energy. 
		Right: for an excitonic insulator, the excitonic mode is overdamped and softens to zero energy at $T_{\mathrm{c}}$. \textbf{g}, Expected Raman spectra at low temperature: due to the coherence factors \textbf{e}, the response at the gap edge is suppressed to zero in the $ac$ scattering geometry, but not in $aa$ (see Methods for details). On heating the features move to lower frequencies (dashed lines).}
	\label{fig:cart}
\end{figure*}

As an example, controlling the bare band gap of an excitonic insulator allows one to explore a range of correlated regimes: from a weakly correlated electon-hole condensate analogous to the Bardeen-Cooper-Schrieffer (BCS) condensate in fermionic superfluids in the negative-gap (semimetallic) regime to a weakly interacting gas of tightly bound excitons in the opposite limit of a gapped (semiconductor) band structure \cite{jerome1967,Ex1968,Transport2017}. In the former case, weakly bound excitons are characterized by size $\xi_{\mathrm{ex}}$ larger than the interparticle distance $l_{\mathrm{eh}}$ and the exciton wavefunctions overlap strongly, while in the latter one, a dilute gas of tightly bound excitons with $\xi_{\mathrm{ex}}\ll l_{\mathrm{eh}}$ exists also above the transition temperature, with their chemical potential going to zero at the transition, in analogy with the Bose-Einstein condensation (BEC). The 'BCS' and 'BEC' regimes are also characterized by a different dynamics of the excitons. In particular, for a semimetallic normal state (as in Fig.\,\ref{fig:cart}d), the exciton, regardless of its energy, may decay into unbound particle-hole pairs (Landau damping), leading to overdamped dynamics.
On the other hand, for a semiconductor-like normal state, the exciton energy is within the direct gap at $T>T_{\mathrm{c}}$, such that energy conservation ensures the undamped dynamics of the excitons. Most interesting is the strongly correlated crossover regime \cite{randeria2014,strinati2018}, where neither of the weakly-coupled bosonic or fermionic descriptions applies and the signatures of excitons above $T_{\mathrm{c}}$ may coexist with the strongly coupled electron-hole plasma.

However, proving the excitonic origin of the phase transition is challenging as it is expected to be obscured by an accompanying structural transition of the crystalline lattice. Formation of a macroscopic excitonic population in a real material may break some of the lattice symmetries: translational one if the excitons are indirect (i.e. have a nonzero center-of-mass momentum) \cite{TS2017} or point-group ones otherwise \cite{Ex1968}. In the latter case, the nontrivial effect on the lattice comes about due to the transformation properties of the particle and hole wavefunctions involved in forming an exciton: if those are distinct, e.g. an {\it s}-like electron is paired with a {\it d}-like hole, the exciton wavefunction would have symmetry lower than the lattice one, even if the relative motion of the particle and hole is in a fully symmetric {\it s}-like state. The discrete nature of the point-group symmetry broken in the excitonic insulator at the transition temperature $T_{\mathrm{c}}$ bears important consequences for its properties: in particular, its excitations are expected to have a finite energy gap, in contrast to the non-equilibrium excitonic condensates, where a continuous $U(1)$ symmetry associated with approximate exciton number conservation is broken resulting in superfluidity and a gapless Bogoliubov-Goldstone mode at $T<T_{\mathrm{c}}$ \cite{snoke2002,deng2010,byrnes2014}.

The above mentioned difficulties are pertinent to the case of Ta$_2$NiSe$_5$, a material showing a semiconducting behavior at low temperatures with a phase transition from high-temperature orthorhombic phase to a monoclinic one at $T_{\mathrm{c}}=328$ K \cite{Structure1986}, breaking two of the mirror symmetries (Fig.\,\ref{fig:cart}, a and b). Symmetry-wise, this transition corresponds to the reduction of point group symmetry from $D_{\mathrm{2h}}$ to $C_{\mathrm{2h}}$ with the order parameter transforming as the $B_{\mathrm{2g}}$ irreducible representation of $D_{\mathrm{2h}}$ ($xz$-like quadrupole). The electronic structure of Ta$_2$NiSe$_5$ has been predicted to have a small or negative (as in Fig.\,\ref{fig:cart}d) direct gap at the Brillouin zone center \cite{Kaneko2012,STM2019}, in agreement with experiments above $T_{\mathrm{c}}$ \cite{Transport2017,IR2017}. The two bands closest to Fermi energy have quasi-1D character and are derived from superpositions of Ta $5d$ and Ni $3d$ orbitals at the multiple sites of the unit cell. Hybridization between them in the $k_{\mathrm{x}}=0$ plane, forbidden by $x\to-x$ symmetry above $T_{\mathrm{c}}$, can serve as the order parameter that is microscopically induced by a condensation of the resulting interband excitons \cite{Theory2013,millis2019}. This points to the quadrupolar character of these excitons.

Experimentally, below $T_{\mathrm{c}}$ formation of a gap has been observed in transport \cite{Transport2017} and optical \cite{IR2017} measurements, and an anomalous dispersion of the hole-like band \cite{ARPES2009} has been taken as an indication for the excitonic character of the transition. 
Above $T_{\mathrm{c}}$, a gap-like spectral weight suppression has been observed in ARPES studies \cite{ARPES2014} suggesting a BEC-like picture for the excitons, while the low values of transport gap suggest otherwise \cite{Transport2017}. 
On the other hand, the changes in spectral and transport properties could also be due to the change of the lattice structure below $T_{\mathrm{c}}$. While Ta$_2$NiSe$_5$ has been actively investigated since then~\cite{ARPES2014,Transport2017,IR2017,IR2018,IR2018a,millis2019}, a structural origin of the transition has not been excluded~\cite{ARPES2020,subedi2020,baldini2020}. Additionally, the origin of the low-temperature spectral gap has not been directly probed: while the valence band dispersion at low temperatures is consistent with a symmetry-breaking hybridization forming below $T_{\mathrm{c}}$~\cite{ARPES2009}, a direct proof of hybridization (regardless of the driving force behind the transition) requires showing the states of two bands being mixed into coherent superpositions of the states at the gap edge as in Fig.\,\ref{fig:cart}d.

The questions above can be addressed directly by Raman spectroscopy that probes the excitations of the system by an inelastic two-photon process. Polarization analysis of the incoming and outgoing photons further enables one to select excitations with a specific symmetry \cite{ovander1960,devereaux2007}. Applied to Ta$_2$NiSe$_5$ above $T_{\mathrm{c}}$, $ac$ polarization geometry (Fig.\,\ref{fig:cart}c, bottom) probes excitations with the symmetry of $ac$-type quadrupole (i.e. $B_{\mathrm{2g}}$, see Methods), the same as that of the order parameter, allowing direct observation of the soft mode expected at a second-order phase transition (Fig.\,\ref{fig:cart}d). Being even in parity, these excitations are invisible in a conventional light absorption experiment due to the dipole selection rules. The character of the soft mode reveals the origin of the transition. In an excitonic transition in a semimetal, critical fluctuations would have a broad relaxational lineshape due to the Landau damping and are enhanced at low frequencies close to $T_{\mathrm{c}}$ (Fig.\,\ref{fig:cart}f, right). On the other hand, in a structural transition driven by an optical phonon, a sharp spectral peak would soften to zero energy at $T_{\mathrm{c}}$ (Fig.\,\ref{fig:cart}f, left). In addition, the structural transition can also be driven by an instability of the acoustic modes (ferroelastic instability \cite{Salje2012}), which would lead to absence of signatures in Raman scattering, as coupling to light vanishes at $q=0$ for acoustic modes \cite{gallais2016}.

Hybridization between bands can be further revealed by studying the contribution of electron-hole excitations to Raman scattering. Above $T_{\mathrm{c}}$, the $ac$ scattering geometry probes the interband transitions (Fig.\,\ref{fig:cart}d) between the valence (red) and conduction (blue) bands that have an $ac$-type quadrupole ($B_{\mathrm{2g}}$) character. The $aa$ scattering geometry, on the other hand, probes fully symmetric excitations (Fig.\,\ref{fig:cart}c, top), and is limited to intraband transitions only. Below $T_{\mathrm{c}}$, hybridization mixes the states of two bands into coherent superpositions (Fig.\,\ref{fig:cart}e). This results in interference effects at the gap edge, in analogy with the effect of coherence factors in a supercondutor \cite{klein1984}. In particular, for $ac$ geometry a destructive interference occurs between transitions from `red' to `blue' states and vice versa resulting in an exact cancellation (Fig.\,\ref{fig:cart}e, on the right). In contrast, for $aa$ geometry, the destructive interference is between two types of interband transitions which do not cancel exactly, as `red' and `blue' bands couple to light differently (Fig.\,\ref{fig:cart}e). This results in the intensity close to the gap edge being strongly suppressed in $ac$ geometry with respect to $aa$. In Fig.\,\ref{fig:cart}g we present the spectra expected in two polarization channels based on a mean-field model of an excitonic insulator (see Methods and Appendix \ref{app:theor} for details). In contrast to the above description, if hybridization is absent and the low-temperature gap is between the conduction and valence bands, the gap edge corresponds to a purely interband transition. The intensity in $ac$ geometry is then expected to be dominant, clearly distinct from the hybridization gap case.

\begin{figure*}
	\centering
	\includegraphics[width=16.5cm]{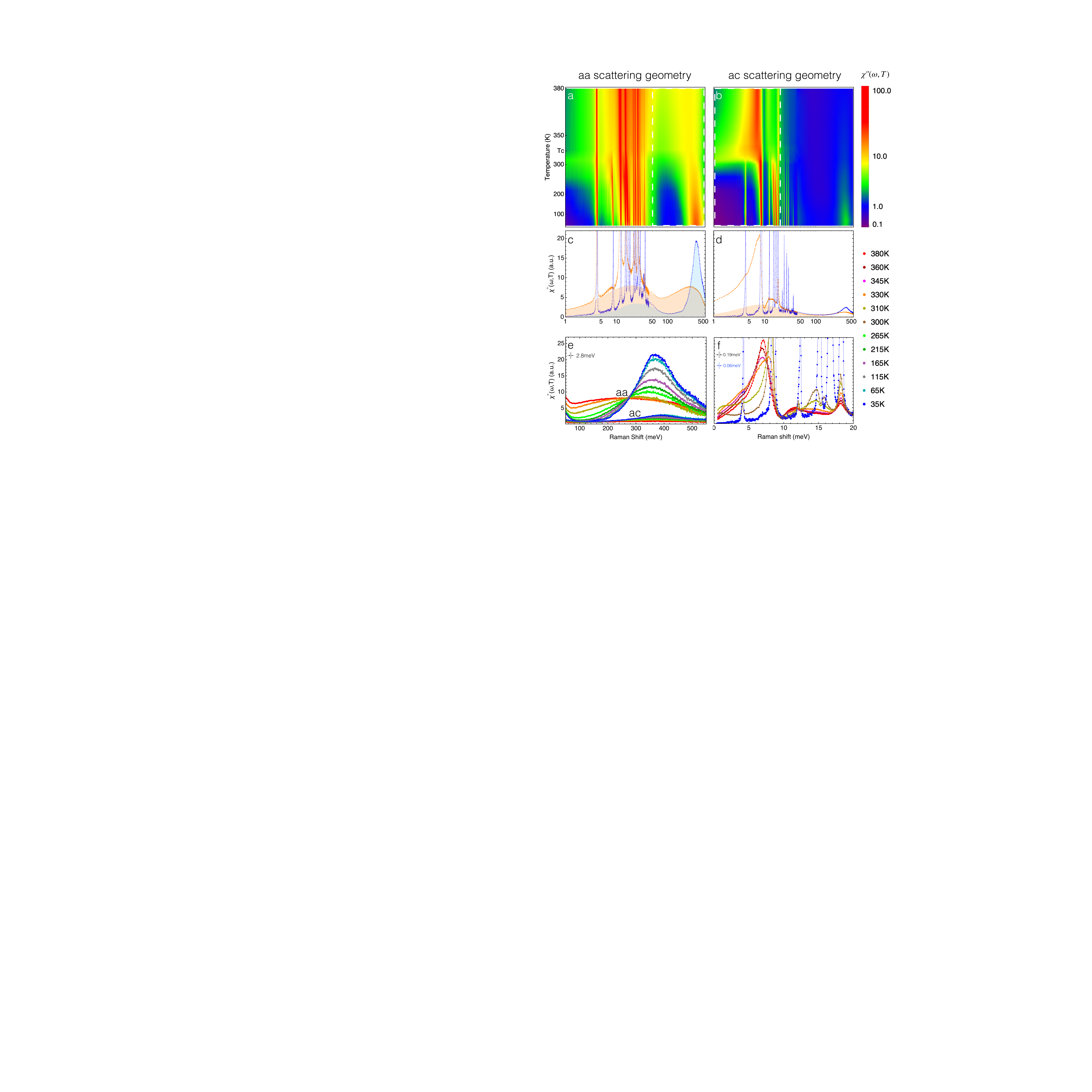}
	\caption{\textbf{Temperature dependence of Raman spectra.} 
		\textbf{a} and \textbf{b}, Raman response $\chi^{\prime\prime}_{\mathrm{aa}}$ and $\chi^{\prime\prime}_{\mathrm{ac}}$ in the $aa$ and $ac$ scattering geometries, respectively. The horizontal frequency (Raman shift) scales are logarithmic. To emphasize the region around $T_{\mathrm{c}}=328\;$K the vertical axes are rescaled by a nonlinear transformation $\arctanh \;[T/T_0]$, where $T_0=390\;$K. 
		$\chi^{\prime\prime}$ values are displayed on a logarithmic false color scale
		% covering more than three decades
		, with the strongest phononic resonances in saturation. % \cite{SM}. 
		In \textbf{b}, a critical enhancement at low frequencies is observed close to $T_{\mathrm{c}}$. Due to broken mirror symmetries below $T_{\mathrm{c}}$, `leakages' of sharp phonon features appear in both \textbf{a} and \textbf{b}. Additionally, a strong peak at $380\;$meV emerges in \textbf{a} with a weaker feature in \textbf{b}. \textbf{c} and \textbf{d}: Details of Raman response  $\chi^{\prime\prime}_{\mathrm{aa}}$ and $\chi^{\prime\prime}_{\mathrm{ac}}$, for $T=330$ K (above $T_{\mathrm{c}}$) and $T=35$ K. Shading highlights the electronic contribution (see also Fig.\,\ref{fig:fanofit}).
		\textbf{e} and \textbf{f}, $\chi^{\prime\prime}_{\mathrm{aa}}$ and $\chi^{\prime\prime}_{\mathrm{ac}}$ in the range enclosed by the dashed white boxes in \textbf{a} and \textbf{b} on a linear frequency scale with the resolution shown in top left corner. In \textbf{e}, $\chi^{\prime\prime}_{\mathrm{aa}}(\omega,T)$ all cross at the 280\,meV isosbestic point, showing intensity transfer on heating from the $380\;$meV peak to the low-temperature gap region. $\chi^{\prime\prime}_{\mathrm{ac}}(\omega,T)$ is shown for intensity comparison. In \textbf{f}, above $T_{\mathrm{c}}$ (red curve) three phonon modes are observed, with the first two showing a pronounced asymmetric Fano lineshape, pointing to an interaction with an electronic continuum. Below $T_{\mathrm{c}}$ the asymmetry becomes less pronounced and additional modes appear due to mutual $aa$-$ac$ `leakage'.}
	\label{fig:data}
\end{figure*}

Here we employ polarization-resolved Raman spectroscopy to prove the excitonic origin of the transition in Ta$_2$NiSe$_5$, as opposed to structural one, and the hybridization nature of the low-temperature gap. We further determine that the resulting excitonic insulator is in the strongly correlated BCS-BEC crossover regime. In particular, close to $T_{\mathrm{c}}$ we observe critical softening of overdamped quadrupolar excitations Fig.\,\ref{fig:data}b,f) that are consistent with excitonic fluctuations in a semimetal. In contrast, we find no softening of the optical phonon modes (Fig.\,\ref{fig:fanofit}, Fig.\,\ref{fig:param}b).
At low temperatures, by comparing the intensities in $aa$ and $ac$ polarization geometries (Fig.\,\ref{fig:data}e) we find direct evidence for hybridization-induced gap and coherent mixing of the two semimetallic bands driven by the excitonic order (as in Fig.\,\ref{fig:cart}d), i.e. spontaneously formed symmetry-breaking hybridization between the bands. With heating, the gap observed in fully symmetric channel fills in, rather than closes, characteristic of strong correlations beyond the mean-field regime. By estimating the exciton coherence length, we find that for Ta$_2$NiSe$_5$ the excitonic condensate lies within the strongly correlated BCS-BEC crossover regime, and argue that the whole body of experimental data for Ta$_2$NiSe$_5$ is consistent with this identification.

\noindent\section*{Results}

\subsection*{Overview}

In Fig.\,\ref{fig:data}a-d we present an overview of the Raman spectra. 
For $ac$ geometry, probing excitations with the symmetry of $ac$-type quadrupole above $T_{\mathrm{c}}$, same as that of the order parameter (Fig.\,\ref{fig:cart}c, bottom), a pronounced enhancement of the low-energy response is observed around $T_{\mathrm{c}}$. 
This is characteristic of a soft mode development near a second order phase transition. 
For $aa$ geometry, the most prominent feature is the redistribution of intensity towards higher energies below $T_{\mathrm{c}}$ with a pronounced gap-like suppression below $380$\,meV; at low energies conventional phonon peaks are observed (Fig.\,\ref{fig:data}c and Appendix, Fig. \ref{fig:aa}).

The presence of a symmetry-breaking phase transition at $T_{\mathrm{c}}$ is evident from the appearance of new sharp optical phonon modes in both geometries. Their appearance is related to the change in selection rules below $T_{\mathrm{c}}$ (see Methods); in what follows, we will call this intensity admixture from an another polarization as ‘leakage’.

This reflects that below $T_{\mathrm{c}}$ the two polarization geometries are no longer orthogonal, thus all excitations may appear in both geometries. 
In Fig.\,\ref{fig:param}a we further quantify this effect by showing the temperature dependence of the integrated intensity of the lowest-energy fully symmetric (above $T_{\mathrm{c}}$) phonon mode in $ac$ scattering geometry, which grows substantially below 328\,K. The parameters of the fully symmetric phonon modes are also consistent with a recent study of Raman scattering in $aa$ geometry \cite{Raman2019}.

\subsection*{Identification of the excitonic soft mode}

\begin{figure}
	\centering
	\includegraphics[width=8cm]{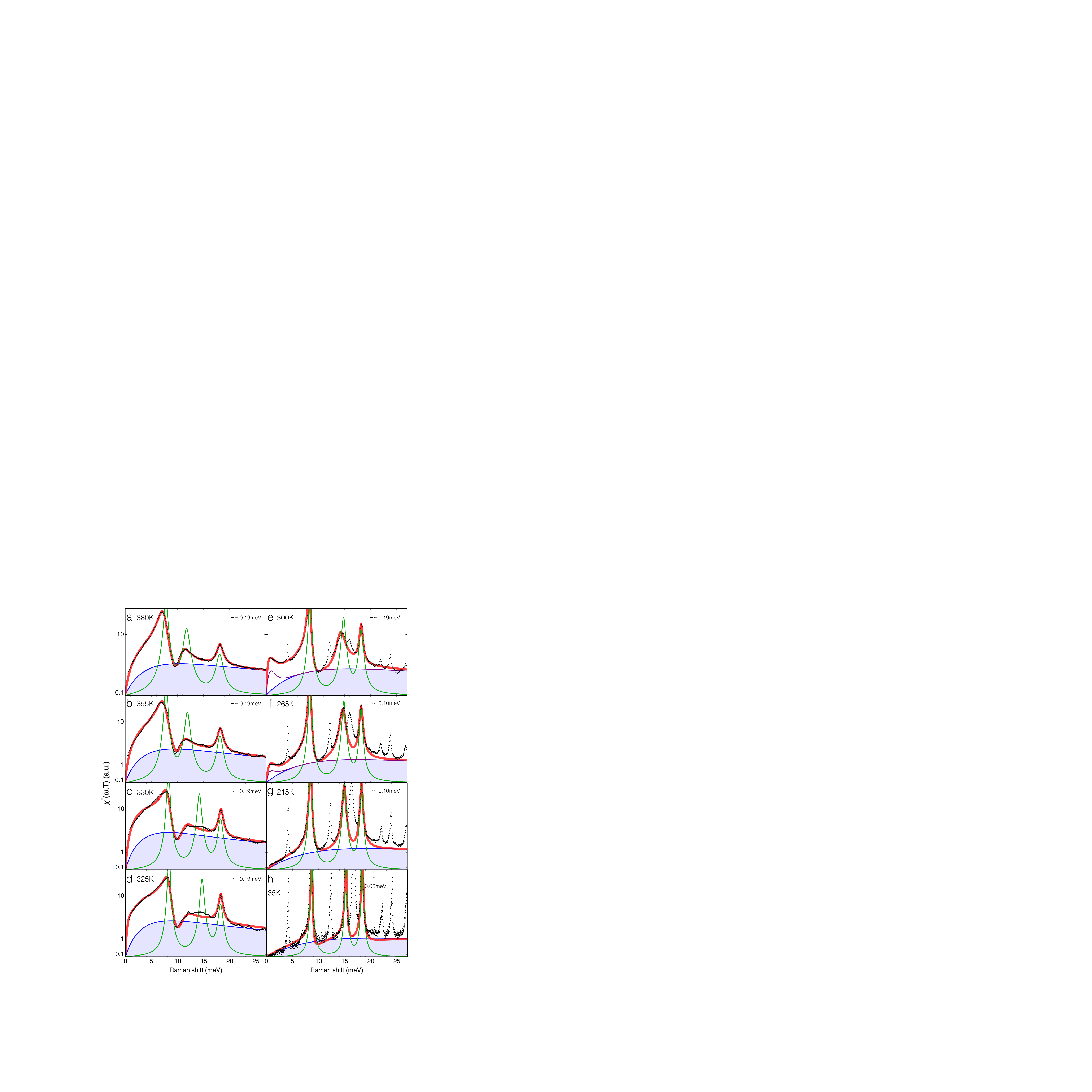}
	\caption{\textbf{Temperature dependence of the low-energy Raman response in the $ac$ scattering geometry.} \textbf{a}-\textbf{h}, show $\chi^{\prime\prime}_{\mathrm{ac}}(\omega)$ data (black dots) for temperatures from $380\;$K to $35\;$K on a semilog scale. Thick red lines show the fits to the data with a generalized Fano model (see Methods) of three phonons coupled to a continuum with a relaxational response $\chi^{\prime\prime}_{\mathrm{cont}}(\omega)$, Eq.\,\eqref{eq:chicont}.
		The deduced bare phonon and continuum responses are shown by green line and blue line with shading, respectively. Full Fano model response (red) does not equal sum of the two due to the presence of interference terms (see Appendix \ref{app:fano}). \textbf{e}-\textbf{f}, 
		Below $T_{\mathrm{c}}$, an additional enhancement at low frequencies is modeled by a low-energy mode, originating from recoil scattering of acoustic phonons on the structural domains (see Methods), interacting with the continuum. Purple line shows its combined response with the continuum. \textbf{g}-\textbf{h}, At lower temperatures $\chi^{\prime\prime}_{\mathrm{cont}}(\omega)$ is suppressed due to gap opening, reducing the asymmetry of the phonon modes. Spectral resolution is shown in top right corner of the panels.
		\label{fig:fanofit}
	}
\end{figure}

We now focus on the low-energy lineshapes in the $ac$ geometry spectra close to the transition temperature, detailed in Fig.\,\ref{fig:data}f and Fig.\,\ref{fig:fanofit}. Three distinct peaks are observed, which correspond to three optical phonons of $B_{\mathrm{2g}}$ symmetry expected from the space group of Ta$_2$NiSe$_5$ ($Cmcm$), ruling the presence of other optical modes out.
The striking feature of the raw data is the notably asymmetric shape of the two lowest-energy
modes, that cannot be described as conventional Lorentzian oscillators.
Instead, the data above $T_{\mathrm{c}}$ is well described by a generalized Fano model (see Methods and Appendix \ref{app:fano} for details), including three phononic oscillators interacting with a continuum of overdamped excitonic excitations described by a purely relaxational response 
\begin{equation}
\chi_{\mathrm{cont}}(\omega) \propto  \frac{1}{- i\omega+\Omega_{\mathrm{e}}(T)},
\label{eq:chicont}
\end{equation}
where its imaginary part $\chi_{\mathrm{cont}}''(\omega)$ exhibits a maximum at $\Omega_{\mathrm{e}}(T)$. This continuum is clearly distinct from the structural phonons and suggests the presence of an overdamped bosonic mode emerging from the electronic system consistent with a Landau-damped exciton in a semimetal (Fig.\,\ref{fig:cart}d). The parameter $\Omega_{\mathrm{e}}(T)$ can then be represented by an analogy with an overdamped oscillator as $\Omega_{\mathrm{e}}(T)\equiv \omega_0^2/\Gamma$, $\omega_0$ being the exciton frequency (which is a collective mode of the semimetal, similar to the collective Cooper pair mode above $T_{\mathrm{c}}$ in a superconductor \cite{larkin2005theory}) and $\Gamma\gg\omega_0$ --- the damping rate. 
This form can be alternatively derived using the time-dependent Ginzburg-Landau equation (see Appendix \ref{app:fano}).
Upon cooling towards $T_{\mathrm{c}}$, $\Omega_{\mathrm{e}}(T)$ linearly decreases as $\Omega_{\mathrm{e}}(T)\propto T- T_{\mathrm{c}}^{el}$ (see Appendix, Fig. \ref{fig:S1}), where $T_{\mathrm{c}}^{el}=137(16)$\,K, consistent with a critical softening of this excitonic collective mode. Just below $T_{\mathrm{c}}$, the $ac$ Raman response is additionally enhanced at the lowest frequencies (Fig.\,\ref{fig:fanofit}e,f). We associate this enhancement with coupling to acoustic $B_{\mathrm{2g}}$ modes with finite momenta, that is mediated by a quasi-periodic pattern of structural domains (which form below $T_{\mathrm{c}}$) that takes the quasimomentum recoil (see Methods for the details).

\begin{figure}
	\centering
	\includegraphics[width=8cm]{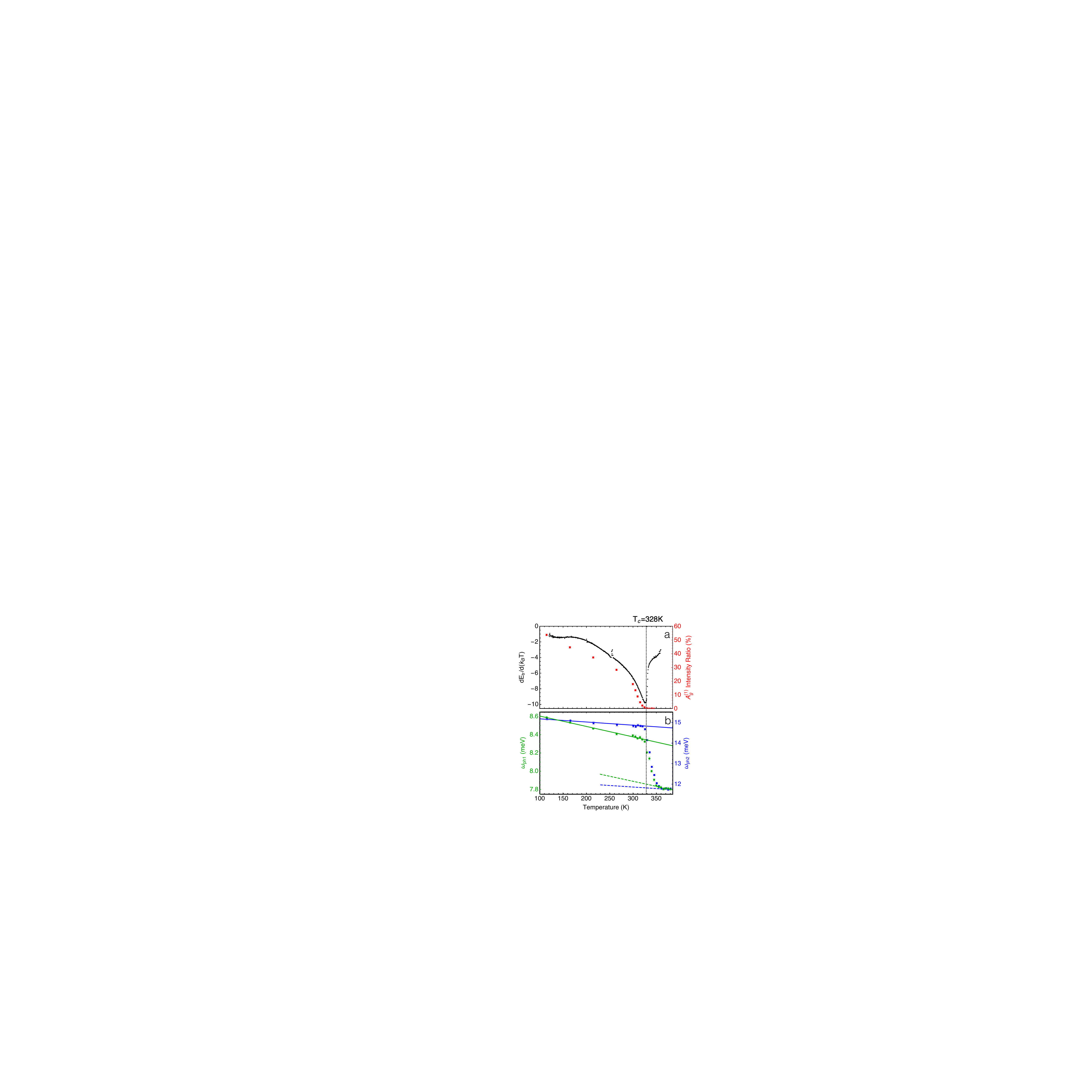}
	\caption{\textbf{Temperature dependence of the phonon parameters.} \textbf{a}, The ratio between the integrated intensity of the lowest-energy $A_{\mathrm{g}}^{(1)}$ phonon mode in the $ac$ and $aa$  scattering geometries. Its appearance below $T_{\mathrm{c}}$ implies the onset of symmetry breaking that mixes the responses in $aa$ and $ac$ geometries. For comparison, the temperature derivative of the transport activation gap $E_{tr}(T) = k_{\mathrm{B}} T \log \left[\frac{R(T)}{R(360\;K)}\right]$ is shown in black, displaying a discontinuity at $T_{\mathrm{c}}$ Error bars are the standard deviation. \textbf{b}, The temperature dependence of the frequency of the two lowest-energy optical phonon modes $\omega_{ph1,2}$ in $ac$ geometry, that exhibit strong coupling with the excitonic continuum. Error bars are the 95\% confidence intervals of the Fano model fit. The phonon frequencies soften on heating consistent with anharmonic decay model (solid lines) below T$_{\mathrm{c}}$ and above $350$\,K (see Appendix \ref{app:fano} for details). The more pronounced change between this regimes is attributed to a nonlinear exciton-phonon coupling, not included in the Fano model (see text).
	}
	\label{fig:param}
\end{figure}

Most importantly, we do not observe phonon softening: instead, the two bare lowest $B_{\mathrm{2g}}$ mode frequencies harden on cooling, Fig.\,\ref{fig:param}b. Consequently, an instability of a zone-center optical phonon \cite{subedi2020} can be ruled out by our results. Away from $T_{\mathrm{c}}$, the hardening on cooling is consistent with the expectations due to lowest-order phonon anharmonicity (lines in Fig.\,\ref{fig:param}b); however, a more pronounced increase is observed close to $T_{\mathrm{c}}$. We attribute this increase to a nonlinear coupling between the excitonic fluctuations and the phonons. In particular, introducing the electronic order parameter $\varphi$, the phonon frequency can be expressed as $\omega_{\mathrm{ph}}(T,\varphi)\approx\omega_{\mathrm{ph}}(T,0)+\alpha_{\mathrm{ph}} \varphi^2 $. For $\alpha_{ph}>0$ (which points to a competition between the electronic and lattice orders), below $T_{\mathrm{c}}$ an additional increase of the phonon frequency is expected. In our results, the increase starts above $T_{\mathrm{c}}$, where $\langle\varphi\rangle=0$, pointing to the influence of the fluctuations beyond the mean-field; in particular, the increase starts at $350$ K suggesting a fluctuation region of around 20 K for the phonon properties.

\begin{figure}
	\centering
	\includegraphics[width=8cm]{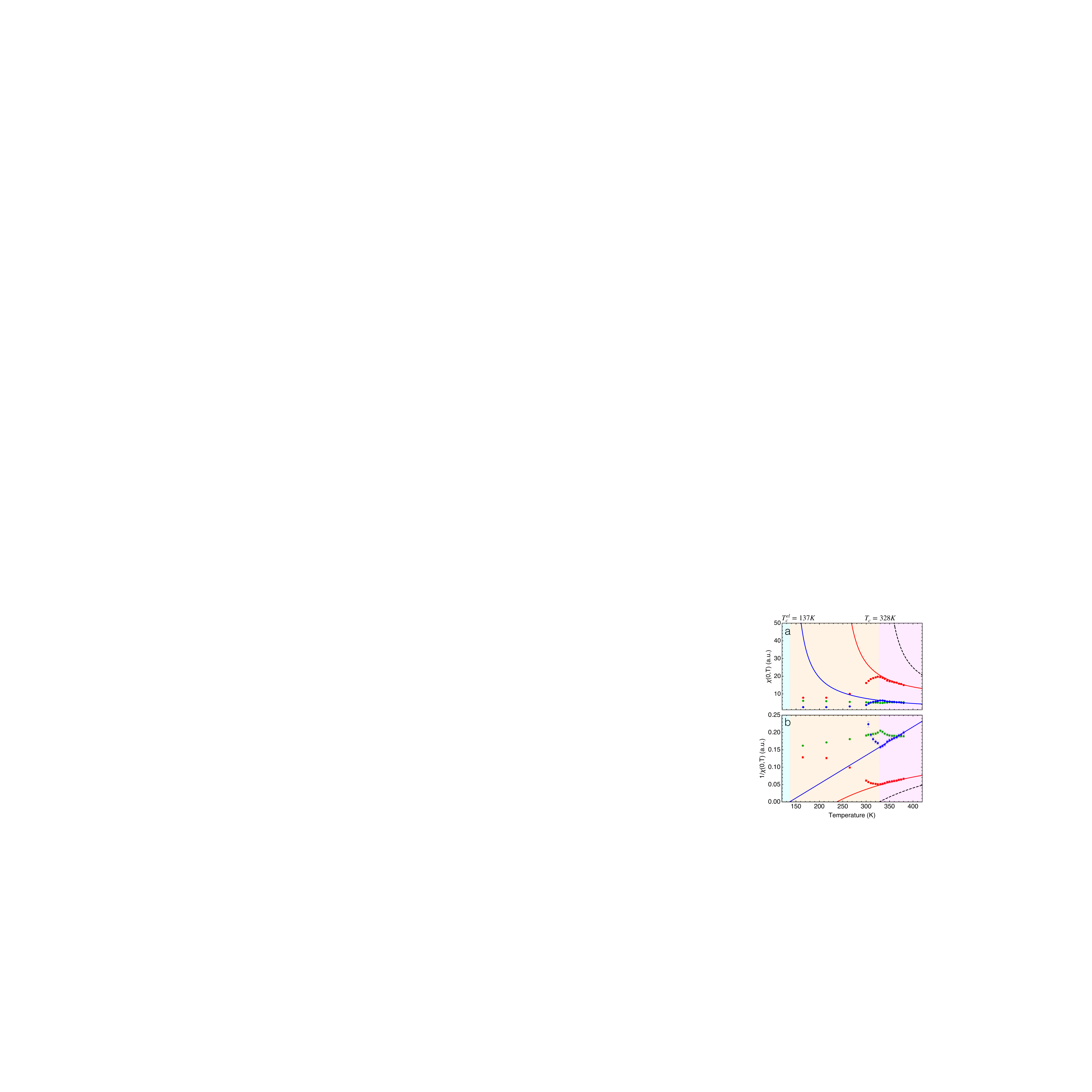}
	\caption{
		\textbf{Static $B_{\mathrm{2g}}$ susceptibility derived from the Fano fits to the Raman data shown in Fig.\,\ref{fig:fanofit}.}
		\textbf{a}, Temperature dependence of the $ac$ quadrupole static susceptibilities $\chi_{\mathrm{ac}}(0,T)$ derived from the decomposition of $\chi_{\mathrm{ac}}''(\omega,T)$ (Fig.\,\ref{fig:fanofit}): excitonic (blue), phononic (green) and combined (red). Above $T_{\mathrm{c}}$ (purple background), while the phononic susceptibility mildly decreases on cooling, the electronic and combined ones grow: lines represent an extrapolation using the Fano model parameters above $350$\,K. The extrapolated electronic susceptibility diverges at $T_{\mathrm{c}}^{el}=137(16)$ K (blue background below). Black dashed line illustrates the total susceptibility including the exciton-strain coupling (see text and Appendix \ref{app:fano}), that diverges at the actual transition temperature $T_{\mathrm{c}}$. Below $T_{\mathrm{c}}$ (pink background), the electronic and combined susceptibilities are suppressed, indicating the formation of the order parameter. 
		\textbf{b}, Same for inverse static susceptibilities $1/\chi_{\mathrm{ac}}(0,T)$. Extrapolated electronic inverse susceptibility shows a linear (Curie-Weiss) behavior vanishing at $T_{\mathrm{c}}^{el}$, while the combined inverse susceptibility is nonlinear and vanishes at a higher temperature, showing that the coupling to otherwise stable (green line) optical phonons can strongly enhance transition temperature. Same is true for coupling to acoustic strain, that can further enhance it to the actually observed $T_{\mathrm{c}}$. Error bars in all panels are the 95\% confidence intervals of the Fano model fit.}
	\label{fig:susc}
\end{figure}

To study the effects of the exciton-phonon coupling on the transition, we analyze the static order parameter susceptibilities (Fig.\,\ref{fig:susc}). From the Fano model, one can deduce the individual susceptibilities of the excitonic continuum, optical phonons as well as the combined one, that includes the effects of the coupling between them. The purely electronic contribution to the inverse susceptibility (Fig.\,\ref{fig:susc}b) $\chi^{-1}_{\mathrm{cont}}(0,T) = \Omega_{\mathrm{e}}(T)/t_{\mathrm{e}}^2$ (from Eq. \eqref{eq:chicont}) follows the Curie-Weiss form $\propto(T-T_{\mathrm{c}}^{el})$ above $T_{\mathrm{c}}$, indicating the softening of $\Omega_{\mathrm{e}}(T)$. Extrapolating this trend to $T<T_{\mathrm{c}}$ suggests that a purely electronic transition would have taken place at $T_{\mathrm{c}}^{el}<T_{\mathrm{c}}$ marked by the divergence of  $\chi_{\mathrm{cont}}(0,T_{\mathrm{c}}^{\mathrm{el}})$. On the other hand, the bare optical phonon susceptibility remains almost constant and is even reduced around $T_{\mathrm{c}}$.

Nevertheless, the coupling between the normal modes of the lattice and the excitonic continuum can increase the transition temperature. Extrapolating the susceptibility of the full Fano model, which includes the exciton-phonon interaction, from above $T_{\mathrm{c}}$ (red line in Fig.\,\ref{fig:susc}) we obtained a transition temperature $T_{\mathrm{c}}^{comb}= 238(18)$ K, larger than $T_{\mathrm{c}}^{el}$ by about 100\,K. However, $T_{\mathrm{c}}^{comb}$ is still smaller then the actual $T_{\mathrm{c}}$, because apart from the optical phonons that we have observed, the coupling to the acoustic $B_{\mathrm{2g}}$ strain modes has also to be considered. Such a coupling has been demonstrated to increase the temperature of an electronic nematic ordering to the actually observed one in iron-based superconductors \cite{chu2012,bohmer2014}. These effects can be understood within the Landau theory, where the electronic order parameter $\varphi$ couples linearly to (optical) phononic and strain order parameters (denoted as $\eta_i$ and $\varepsilon_{\mathrm{ac}}$, respectively) of the same symmetry resulting in the free energy expansion
\begin{equation}
\begin{gathered}
F[\varphi,\eta_i,\varepsilon_{\mathrm{ac}},T] = \chi_{\mathrm{cont}}^{-1}(0,T)\frac{\varphi^2}{2}+ \sum_{i=1}^3\chi_{\mathrm{opt},i}^{-1}(0,T)\frac{\eta_i^2}{2}
\\
+\lambda_{\mathrm{opt},i} \varphi \eta_i 
+
\kappa\frac{\varepsilon_{\mathrm{ac}}^2}{2}+\lambda_{\mathrm{ac}} \varphi \varepsilon_{\mathrm{ac}}
+O(\varphi^4),
\label{eq:LandF}
\end{gathered}
\end{equation}
where $\kappa$ is the $B_{\mathrm{2g}}$ shear modulus. Minimizing the quadratic term one obtains the condition for the transition temperature $\chi_{\mathrm{cont}}^{-1}(0,T_{\mathrm{c}})-\lambda^2/\kappa-\sum_{i=1}^3\lambda^2_{\mathrm{opt},i}\chi_{\mathrm{opt},i}(0,T_{\mathrm{c}}) =0$, as opposed to $\chi_{\mathrm{cont}}^{-1}(0,T_{\mathrm{c}}^{el})=0$ in the purely electronic case, leading to an enhanced transition temperature $T_{\mathrm{c}}$. 
In Fig.\,\ref{fig:susc} we illustrate this effect (dashed black line) by choosing the value of the interaction with strain such that the total susceptibility diverges at $T_{\mathrm{c}}$. We find the value of the exciton-strain coupling obtained this way to be consistent with a recent study of the sound velocity renormalization above $T_{\mathrm{c}}$ \cite{XRay2018} (see Appendix \ref{app:acous}), and the Raman response from finite-momentum acoustic phonons below $T_{\mathrm{c}}$ in Fig.\,\ref{fig:fanofit}e,f.

The analysis above leads us to the following conclusions: (a) the excitonic mode exhibits a strong tendency to soften, suggesting a purely electronic transition temperature of $T_{\mathrm{c}}^{el}=137(16)$\,K; and (b) the transition temperature is boosted to the observed $T_{\mathrm{c}}=328$ K due to the coupling to noncritical optical phonon modes and acoustic strain.

\subsection*{Coherent factors in the excitonic insulator}

Having established the excitonic origin of the transition in Ta$_2$NiSe$_5$ we now demonstrate that an interband hybridization emerges in the excitonic insulator phase. At low temperatures, an intense peak at about $380$\,meV in the $aa$ geometry emerges with a much subtler feature in the $ac$  geometry (Fig.\,\ref{fig:data}e). This stark contrast cannot be attributed to the difference of the Raman coupling in two geometries: while the intensity in the $ac$ geometry is indeed weaker than in the $aa$ one at all temperatures, the evolution of the intensity in $aa$ geometry at $T<T_{\mathrm{c}}$ is much more pronounced, allowing to relate this effect to the symmetry breaking (see also Appendix \ref{app:highen}). Moreover, a substantial part of the signal observed in $ac$ geometry at low temperatures can be attributed to the change in selection rules below $T_{\mathrm{c}}$, making the $ac$ intensity actually related to the interband transition even smaller (see Appendix Fig. \ref{HE}). Both band structure calculations \cite{Theory2013} and experiments \cite{ARPES2018TR,fukutani2019,chen2020} suggest that only two bands are mostly affected by this symmetry breaking, allowing us to limit ourselves to the simplest excitonic insulator model of Fig.\,\ref{fig:cart}. The strong temperature dependence of this feature below $T_{\mathrm{c}}$ (also discussed below) also precludes its attribution to transitions away from $k_x=0$ plane, where hybridization can be present already above $T_{\mathrm{c}}$. For a hybridization-induced gap in a semimetal, a suppression of interband transitions at the gap edge in the $ac$ geometry is indeed expected (Fig.\,\ref{fig:cart}e). Within a simplified two-band model of the excitonic insulator (Fig.\,\ref{fig:cart}g), a divergence of the intensity at the gap edge occurs in $aa$ geometry, but not in the $ac$ one, where interference effects lead to an exact cancellation of the intensity at the gap edge. These predictions are in agreement with the pronounced peak being observed in the $aa$ geometry only (which is further exacerbated if the leakage from $aa$ to $ac$ geometry is accounted for, see Appendix Fig. \ref{HE}).
On the other hand, in the case of a semiconductor with a large direct gap, such that hybridization effects can be ignored, the opposite is true: the interband transition would have predominantly $ac$ quadrupolar character allowing its observation only in $ac$ geometry (see Methods and Appendix \ref{app:theor} for details), in stark contrast to the data in Fig.\,\ref{fig:data}e. 
Thus, the observation of an intense peak emerging at low temperatures in $aa$ geometry clearly favors a semimetallic band structure with a hybridization-induced gap.

Now we discuss the temperature dependence of the high-energy Raman response in $aa$ geometry (Fig.\,\ref{fig:data}e). Within the mean-field theory, the energy of the peak is related to the amplitude of the order parameter, which is expected to diminish upon heating towards $T_{\mathrm{c}}$ (Fig.\,\ref{fig:cart}g), and zero intensity is expected within the hybridization gap. We have observed a pronounced redistribution of intensity in a broad energy range, with intensity appearing at the lowest energies at temperatures significantly smaller than $T_{\mathrm{c}}$. Such behavior appears inconsistent with the mean-field expectation, pointing to a role of correlations beyond the mean-field picture. This is further emphasized by all the curves crossing at 280\,meV - the so-called 'isosbestic point' \cite{freericks2001,devereaux2007,greger2013}, characteristic of strongly correlated systems (see also the discussion of BCS-BEC crossover effects and pseudogap below). It was previously observed in materials where spectral gap has a many-body origin, such as SmB$_6$~\cite{Nyhus1996} or cuprates~\cite{guyard2008}. While band gap renormalization in semiconductors can be also related to electron-phonon interactions \cite{giustino2010}, the broadening of the gap edge features is expected to be rather small in that case; on the contrary, the spectral features we have observed quickly become broader than their energy on heating (Fig.\,\ref{fig:data}e). Furthermore, the renormalization effects are expected to diminish with phonon energy \cite{karsai2018} and in our case all phonon energies are below $50$ meV, suggesting the scale of the electron-phonon effects to remain below 100 meV.

\noindent\section*{Discussion}

Our results unambiguously point to the excitonic physics playing a crucial role in Ta$_2$NiSe$_5$; we further found evidence for strong correlation effects beyond the mean-field expectations. The important role of correlations in the formation of excitonic order in Ta$_2$NiSe$_5$ is further corroborated by the large ratio $\frac{2\Delta}{k_{\mathrm{B}} T_{\mathrm{c}}}\approx13$, $k_{\mathrm{B}}$ being the Boltzmann constant ($2\Delta \approx 380$\,meV taking the energy of the spectral peak in Fig.\,\ref{fig:data}e as an estimate for the order parameter value), well beyond the BCS prediction, but consistent with the suppression of $T_{\mathrm{c}}$ occurring in BCS-BEC crossover~\cite{randeria2014}. Moreover, we can estimate the coherence length of the excitonic order using the bare electronic transition temperature with the BCS expression $\xi_{\mathrm{ex}} = \frac{\hbar v_{\mathrm{F}}}{1.76\pi k_{\mathrm{B}} T_{\mathrm{c}}^{el}}$. As the system is highly anisotropic, we use the value of $v_{\mathrm{F}}$ along the most dispersive $a$ direction, where the exciton size is expected to be largest. This yields $\xi_{\mathrm{ex}} = 32$\,\AA, while the distance between particles can be estimated in a quasi-1D system from $l_{\mathrm{eh}}=\frac{\pi}{2k_{\mathrm{F}}}\approx 16$\,\AA \, ($k_{\mathrm{F}}\sim 0.1$\,\AA$^{-1}$,$m^*\approx0.37 m_0$~\cite{ARPES2018TR}).
The estimate places the excitons in Ta$_2$NiSe$_5$ in the correlated BCS-BEC crossover regime $l_{\mathrm{eh}}\sim\xi_{\mathrm{ex}}$. 

Identification of Ta$_2$NiSe$_5$ as a correlated excitonic insulator finally allows to clarify the character of the electronic structure above $T_{\mathrm{c}}$. Semimetallic character is suggested by the absence of a discernible gap in spectroscopy (Fig.\,\ref{fig:data}e and Refs. \cite{IR2017,ARPES2018TR}) and the overdamped lineshape of the excitonic fluctuations (Fig.\,\ref{fig:fanofit}). A direct semiconducting gap at $T>T_{\mathrm{c}}$ would further contradict the absence of a noticeable peak in $ac$ Raman intensity (Fig.\,\ref{fig:data}e) at low temperatures that requires strong mixing of the bands as in Fig.\,\ref{fig:cart}e. On the other hand, an apparent gap feature has been observed in ARPES studies above $T_{\mathrm{c}}$~\cite{ARPES2014}. It is known however, that strong fluctuations of a particle-hole order, such as a charge density wave, result in a pseudogap~\cite{lee1973,schmalian1998,sadovskii2001} in a metallic system - a suppression of the density of states already above $T_{\mathrm{c}}$, which has been experimentally observed close to charge- or spin-density wave transitions~\cite{borisenko2008,boschini2020}. In all of these cases, the pseudogap opens above the ordering transition due to the thermal order parameter fluctuations, regardless of the microscopic mechanism of the transition. Strong excitonic fluctuations in the BCS-BEC crossover regime may be expected to lead to similar effects, which explains the spectral weight suppression observed in ARPES \cite{ARPES2014}. Moreover, as the density of states in the pseudogap is finite \cite{lee1973}, it can be reconciled with both the zero-gap behavior of the high-temperature transport \cite{Transport2017} as well as the overdamped character of the excitonic mode (Fig.\,\ref{fig:fanofit}).

In conclusion, by using polarization-resolved Raman spectroscopy, we have directly revealed the excitonic fluctuations driving a phase transition in Ta$_2$NiSe$_5$, shown the coherence of the low-temperature insulator by revealing the coherent band superpositions, and identified the fingerprints of strong electronic correlations, similar to the ones occuring in the BCS-BEC crossover. The strongly correlated excitonic insulator nature of Ta$_2$NiSe$_5$ accommodates for its anomalous features and provides a unified view of its electronic structure, consistent with the previous experiments \cite{ARPES2009,ARPES2014,IR2017}. Furthermore, as the band structure of this material can be tuned by chemical substitution or pressure \cite{Transport2017}, this opens perspectives for the exploration of different correlation regimes in excitonic insulators.

%\newpage

\noindent\section*{Methods}

\vspace{-2mm}
\noindent \textbf{Sample preparation.}
Single crystals of Ta$_2$NiSe$_5$ were grown using the chemical vapor transport (CVT) method. 
Elemental powders of tantalum, nickel and selenium were mixed with a stoichiometric ratio and then sealed in an evacuated quartz ampoule  with a small amount of iodine as the transport agent. The mixture was placed in the hot end of the ampoule ($\sim$950$^{\circ}$C) under a temperature gradient of about 10$^{\circ}$C cm$^{-1}$. 
After about a week mm-sized needle-like single crystals were found at the cold end of the ampoule. 
These crystals are shiny and cleave easily. 
We have used x-ray diffraction and Electron Dispersive X-ray Spectroscopy (EDS) to verify the exact composition of the crystals and their uniformity. 

\vspace{2mm}
\noindent \textbf{Resistance Measurement.}
The resistance along the a-axis (direction along the Ta/Ni chains) of a very thin needle-like single crystal was measured in a four-probe configuration using a Quantum Design DynaCool PPMS system 
(see Appendix \ref{app:exp} for details).

\vspace{2mm}
\noindent \textbf{Polarization-resolved Raman measurements.}
The samples used for Raman measurement were cleaved to expose (010) crystallographic plane ($ac$ plane). 
The cleaved surface was then examined under a Nomarski microscope to find a strain-free area.

Raman measurements were performed in a quasi-back-scattering geometry from the samples placed in a continuous helium-gas-flow cryostat. 
The 647\,nm line from a Kr$^+$ ion laser was used for excitation. 
Incident light was focused into a 50$\times$100\,$\mu$m$^{2}$ spot of freshly cleaved crystal surface. For data taken below 310\,K, laser power of 8\,mW was used. To reach temperature above 310\,K, we kept the environmental temperature at 295\,K and increased laser power to reach higher sample temperature in the excitation spot. The reported temperatures were corrected for laser heating, which was determined to be 1.29$\pm$0.17\,K\,mW$^{-1}$ from the Stokes/anti-Stokes intensity measurement consistent with detailed balance (see Appendix \ref{app:temp} for details). We employed a custom triple-grating spectrometer with a liquid-nitrogen-cooled charge-coupled device detector for analysis and acquisition of the Raman signal. 
All the data were corrected for the spectral response of the spectrometer and the CCD detector. For low-frequency Raman shift data below 50\,meV, the spectral resolution was set to 0.06\,meV for data taken below 100\,K; to 0.10\,meV for data taken from 100\,K to 295\,K; and to 0.19\,meV for data taken above 295\,K.
For high-frequency spectral features above 50\,meV about 2.5\,meV resolution was used.  

Two incident/scattered light polarization configurations were employed to probe excitations in different symmetry channels. 
The relationship between the scattering geometries and the symmetry channels is given in Table I.

The Raman response function $\chi''(\omega,T)$ was calculated from the measured Raman intensity $I(\omega,T)$ by the relation $I(\omega,T)=[1+n(\omega,T)]\chi''(\omega,T)$, where $n(\omega,T)$ is the Bose factor, $\omega$ is Raman shift energy, and $T$ is temperature.

\vspace{2mm}
\noindent \textbf{Fitting model for the Raman susceptibility.}
We adapt an extended Fano model~\cite{klein1983} comprising three phonon modes interacting with an electronic continuum to fit the $ac$-polarization Raman susceptibility data above $T_{\mathrm{c}}$, see Fig.\,\ref{fig:data}f and Fig.\,\ref{fig:fanofit} of the main text. 
It was found that the electronic continuum is best described by a simple relaxational response $\chi^{\prime\prime}_{\mathrm{cont}}(\omega) = \frac{t_{\mathrm{e}}^2 \omega}{\omega^2 + \Omega_{\mathrm{e}}^2}$, 
corresponding to the overdamped limit of the Drude-Lorentz model (see Eq.\,(1) in the main text). 
Here $t_{\mathrm{e}}$ is the Raman coupling vertex to electronic response and 
$\Omega_{\mathrm{e}}\equiv\frac{\omega_0^2}{\Gamma}$ is a single fit parameter representing ratio of the square of electronic excitation frequency to its relaxation rate.

The resulting purely phononic and electronic parts of the Raman susceptibility are shown in Fig.\,\ref{fig:fanofit} in green and blue lines correspondingly. The temperature dependencies of the frequencies of the lowest phonons are shown in Fig.\,\ref{fig:param}b. The electronic part of the inverse susceptibility (Fig.\,\ref{fig:susc}b, red points) is, on the other hand, proportional to the electronic oscillator parameter $\Omega_{\mathrm{e}}(T)$ (see Eq. \eqref{eq:chicont}). The static susceptibilities shown in Fig.\,\ref{fig:susc} are obtained by taking the Fano model susceptibility at $\omega=0$. As the model susceptibility is an analytic function and matches the raw data well, the result is equivalent to that of a Kramers-Kronig transformation.

By analyzing the data just below $T_{\mathrm{c}}$ (Fig.\,\ref{fig:fanofit} e,f), we have found an additional enhancement at low frequencies can not be described with the same model as the characteristic continuum energy $\Omega_{\mathrm{e}}(T)$ only grows below $T_{\mathrm{c}}$ (see Appendix, Fig. \ref{fig:S1}) and the resulting continuum intensity at low energies only weakens (Fig.\,\ref{fig:fanofit} d-g). This suggests the presence of an additional source of low-frequency response.
The appearance of a low-energy mode can be appreciated from the formation of domains observed in transmission electron microscopy (see Appendix \ref{app:fano} and Ref. \cite{Structure1986}). In particular, the domains have an elongated needle-like shape along the $a$ axis, with their size along the $c$ axis on the order of $\bar{d} = 200$\AA. Acoustic phonon, scattering on this quasiperiodic structure, can take a recoil momentum of the order $2\pi/\bar{d}$; this way a finite $q_{\mathrm{d}}=\pm2\pi/\bar{d}$ phonon can contribute to the Raman intensity (where the momentum transfer is approximately zero). At finite $q$, introducing the acoustic phonon coordinate $x_q = \sqrt{q} u_{\mathrm{q}}$, where $u_{\mathrm{q}}$ - is the Fourier transform of the atomic displacement amplitude, one obtains that the Raman vertex $t_{\mathrm{s}}=\tau_{\mathrm{s}}\sqrt{q}$ and the coupling to the excitonic mode $v_{\mathrm{s}}=\beta_{\mathrm{s}}\sqrt{q}$ of the acoustic phonon are both proportional to $\sqrt{q}$. Both vanish at ${\bf q}=0$ and thus the acoustic modes can not be observed in the absence of domains. The energy of the relevant $B_{\mathrm{2g}}$ acoustic mode is $\omega_{\mathrm{s}}(q) = c_{\mathrm{s}} q$, where $c_{\mathrm{s}}$ is the sound velocity, that can be deduced from the X-ray measurements \cite{XRay2018} to be $c_{\mathrm{s}} \approx 30$\,meV\,\AA\, away from $T_{\mathrm{c}}$. Note that the coupling to the soft excitonic mode results in the renormalization of the observed $c_{\mathrm{s}}$ close to $T_{\mathrm{c}}$; in Appendix \ref{app:acous} it is shown that the softening of $c_{\mathrm{s}}$ close to $T_{\mathrm{c}}$ can be fully attributed to this effect. To fit the data we further fix the ratio $\beta_{\mathrm{s}}^2/c_{\mathrm{s}}$ such that the total static susceptibility (Eq. 1) diverges at $T_{\mathrm{c}}$ and assume that the domain size $i$ in units of the crystal unit cell constant $c$ follows the Poisson distribution. In addition to that, we allow for an intrinsic linewidth of the phonons at finite $q$: $\omega_{\mathrm{s}}\to\omega_{\mathrm{s}}-i r_{\mathrm{s}}q$. Thus, the parameters $\tau_{\mathrm{s}}$ and $r_{\mathrm{s}}$ remain the only fit parameters. Additional details can be found in Appendix \ref{app:exp}.

\vspace{2mm}
\noindent \textbf{Effect of coherence factors on low-temperature Raman scattering.}
To describe an excitonic insulator at low temperature we use a two-band mean field model (see Appendix \ref{app:theor} for details):  $\hat{H}_{\mathrm{MF}} = \sum_{\bf p} \varepsilon_{\mathrm{c}}({\bf p})\hat{c}^\dagger_{\mathrm{c}}({\bf p})\hat{c}_{\mathrm{c}}({\bf p})+\varepsilon_{\mathrm{v}}({\bf p})\hat{c}^\dagger_{\mathrm{v}}({\bf p})\hat{c}_{\mathrm{v}}({\bf p})+W(\hat{c}^\dagger_{\mathrm{c}}({\bf p})\hat{c}_{\mathrm{v}}({\bf p})+\hat{c}^\dagger_{\mathrm{v}}({\bf p})\hat{c}_{\mathrm{c}}({\bf p}))$, where $\varepsilon_{c(v)}({\bf p})$ is the dispersion of the conduction (valence) band and $W$ - is the excitonic order parameter (spin index is omitted for brevity). The order parameter mixes two bands such that the eigenstates are superpositions of band states with coherent factors, analogous to the BCS ones. To calculate the Raman intensity, we use the following form of the electron-light vertex, consistent with the system's symmetry above $T_{\mathrm{c}}$: $
\hat{R}_{\mathrm{aa}} = \sum_p g_{\mathrm{c}} \hat{c}^\dagger_p \hat{c}_p+g_{\mathrm{v}} \hat{v}^\dagger_p \hat{v}_p
;\;
\hat{R}_{\mathrm{ac}} = g_{\mathrm{ac}} \sum_p \hat{c}^\dagger_p \hat{v}_p+  \hat{v}^\dagger_p \hat{c}_p,
$
where $\hat{R}_{\mathrm{aa}}$ is the vertex for $aa$ polarization geometry and $\hat{R}_{\mathrm{ac}}$ - for $ac$ geometry, and $g_{\mathrm{c/v}},g_{\mathrm{ac}}$ are coupling constants.

For the semimetallic case one can linearize the dispersion $\varepsilon_{c,v}({\bf p})$ around the Fermi surface. The resulting Raman intensity obtained using Fermi's Golden Rule is:
\begin{equation}
I_{R_{\mathrm{ac}}} \sim  \frac{4 \pi g_{\mathrm{ac}}^2 \nu_0 \sqrt{\omega^2/4-W^2}}{\omega};
\;
I_{R_{\mathrm{aa}}} \sim  \frac{\pi(g_{\mathrm{c}}-g_{\mathrm{v}})^2 \nu_0 W^2}{\omega\sqrt{\omega^2/4-W^2}},
\end{equation}
for $\omega>2W$ and zero otherwise, where $\nu_0$ is the density of states. One can see that at the gap edge, the $ac$ intensity vanishes due to the vanishing matrix element of the transition, as is shown in Fig.\,\ref{fig:cart}e, while the $aa$ diverges as a square-root, similar to the case of a BCS superconductor \cite{klein1984}, as is shown in Fig.\,\ref{fig:cart}g.

For the semiconductor-like 1D band structure with a direct gap $2E_0$, on the other hand, we find that the interference effects are weakened such that a singularity at the gap edge is present in $aa$ and $ac$ scattering (See Appendix \ref{app:theor} for details).

\section*{Data availability}
\vspace{-2mm}
\noindent
The data that support the plots within this paper and other findings of this study are available from the authors upon reasonable request.

\section*{Code availability}
\vspace{-2mm}
\noindent
Wolfram Mathematica codes for data analyses are available from one of the contributing authors (M.Y.) upon reasonable request. 

\section*{Acknowledgments}
\vspace{-2mm}
\noindent
We thank K. Haule for discussions.

\noindent \textbf{Funding:}
The spectroscopic work conducted at Rutgers was supported by NSF Grant No. DMR-1709161 (M.Y. and G.B). 
%and DMR-1709229 (K.H). 
P.A.V. acknowledges the Postdoctoral Fellowship support from the Rutgers University Center for Materials Theory. 
The sample growth and characterization work conducted at the Technion was supported by the Israel Science Foundation Grant No. 320/17 (H.L., I.F. and A.K.). H.L. was supported in part by a PBC fellowship of the Israel Council for Higher Education. 
The work at NICPB was supported by the European Research Council (ERC) under Grant Agreement No. 885413. 

\noindent \textbf{Author contributions:} 
P.A.V and M.Y. contributed equally to this project. 
G.B. conceived and supervised the project. 
G.B. and M.Y. designed the experiment. 
I.F, H.L. and A.K. grew and characterized the crystals. 
M.Y. performed the spectroscopic experiments.
M.Y and P.A.V. performed the data analysis and prepared the figures. 
%P.A.V. and K.H. developed theoretical formalism and performed calculations. 
%P.A.V., M.Y., A.K. and G.B. contributed to discussions and writing the manuscript. 
All the authors contributed to discussions and writing the manuscript.

\vspace{2mm}
\noindent \textbf{Competing interests:} 
The authors declare no competing interests.

\appendix

\section{Experimental details}
\label{app:exp}

\subsection{Resistance Measurement}
\begin{figure}[h!]
	\centering
	\includegraphics[width=0.5\textwidth]{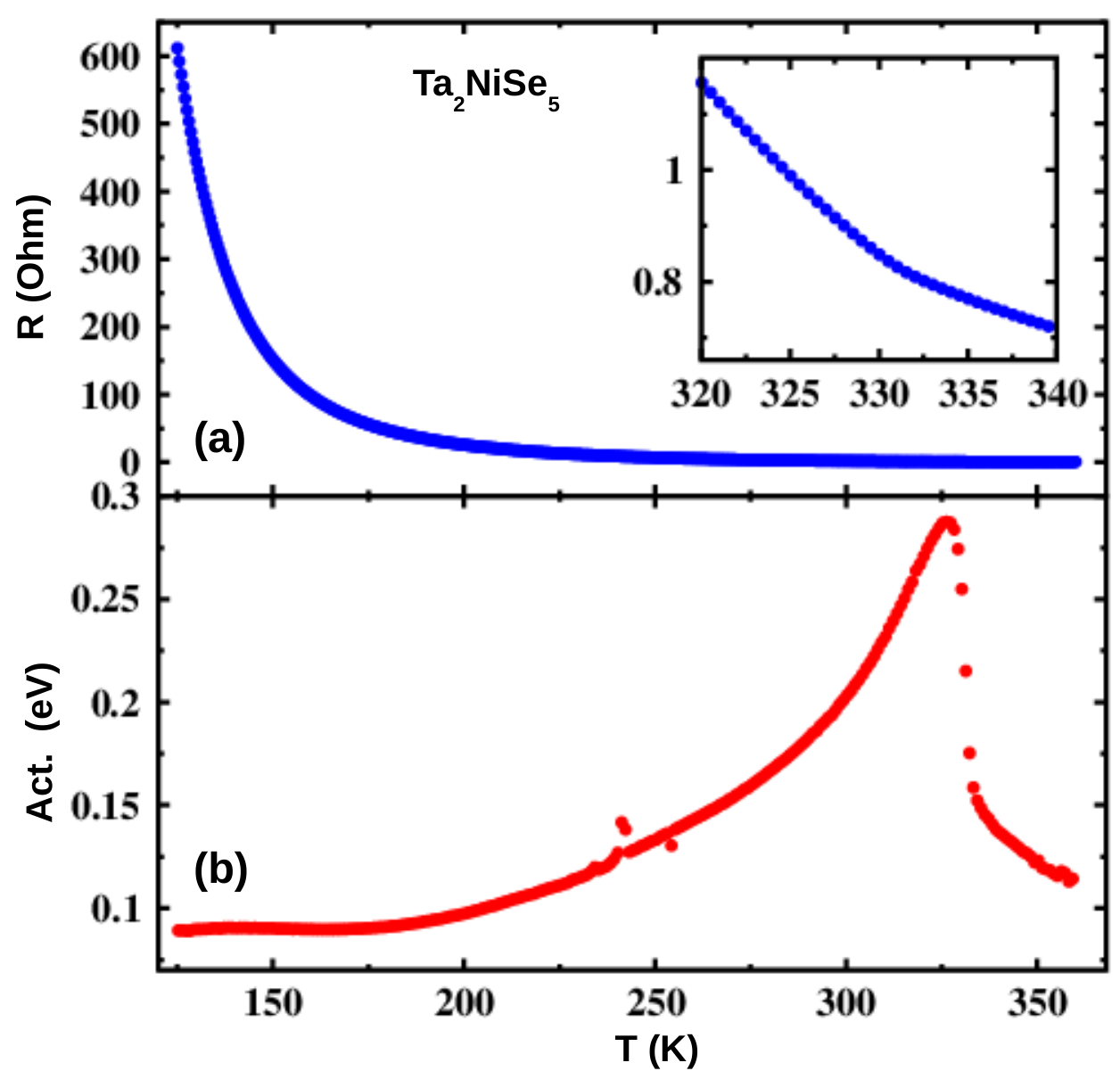}
	\caption{ Resistance (a) and activation energy (b) plot of Ta$_2$NiSe$_5$. Inset of (a) shows a small break in the resistance around the transition temperature 327 K.}
	\label{res}
\end{figure}

Fig \ref{res}(a) shows the resistance as a function of the temperature.  We find a small break in the resistivity at the transition temperature $T_{\mathrm{c}}$ $\simeq$ 327 K, shown in the inset. This small change appears  more prominent in Fig \ref{res}(b), where the activation energy (- k$_B$ T$^2$ $\frac{\partial \ln(R (T))}{\partial T}$) is presented. This result is consistent with the previous report \cite{Transport2017} within the accuracy of the measurement.

\subsection{Heating rate estimation using Stokes/Anti-Stokes analysis}
\label{app:temp} 

In thermal equilibrium, the Stokes (absorption) scattering cross section $I_{\mathrm{S}}$ and Anti-Stokes (emission) cross section $I_{\mathrm{AS}}$ are related by the detailed balance principle~\cite{Hayes2004}
\begin{equation}
n I_{\mathrm{S}}(\omega)=(n+1)I_{\mathrm{AS}}(\omega)~,
\label{eq:SAS}
\end{equation}
in which $n$ stands for the Bose factor
\begin{equation}
n(\omega,T)=\frac{1}{\exp(\hbar\omega/k_{\mathrm{B}}T)-1}~,
\label{eq:n}
\end{equation}
where $\hbar$ is the reduced Planck's constant, $\omega$ is frequency, $k_{\mathrm{B}}$ is the Boltzmann's constant, and $T$ is the temperature in the laser spot. Using Eq.~(\ref{eq:SAS}-\ref{eq:n}), the temperature can be obtained as 
\begin{equation}
T=\frac{\hbar\omega}{k_{\mathrm{B}} \log(I_{\mathrm{S}}(\omega)/I_{\mathrm{AS}}(\omega)}~.
\label{eq:T}
\end{equation}

\begin{figure}[h!]
	\includegraphics[width=0.45\textwidth]{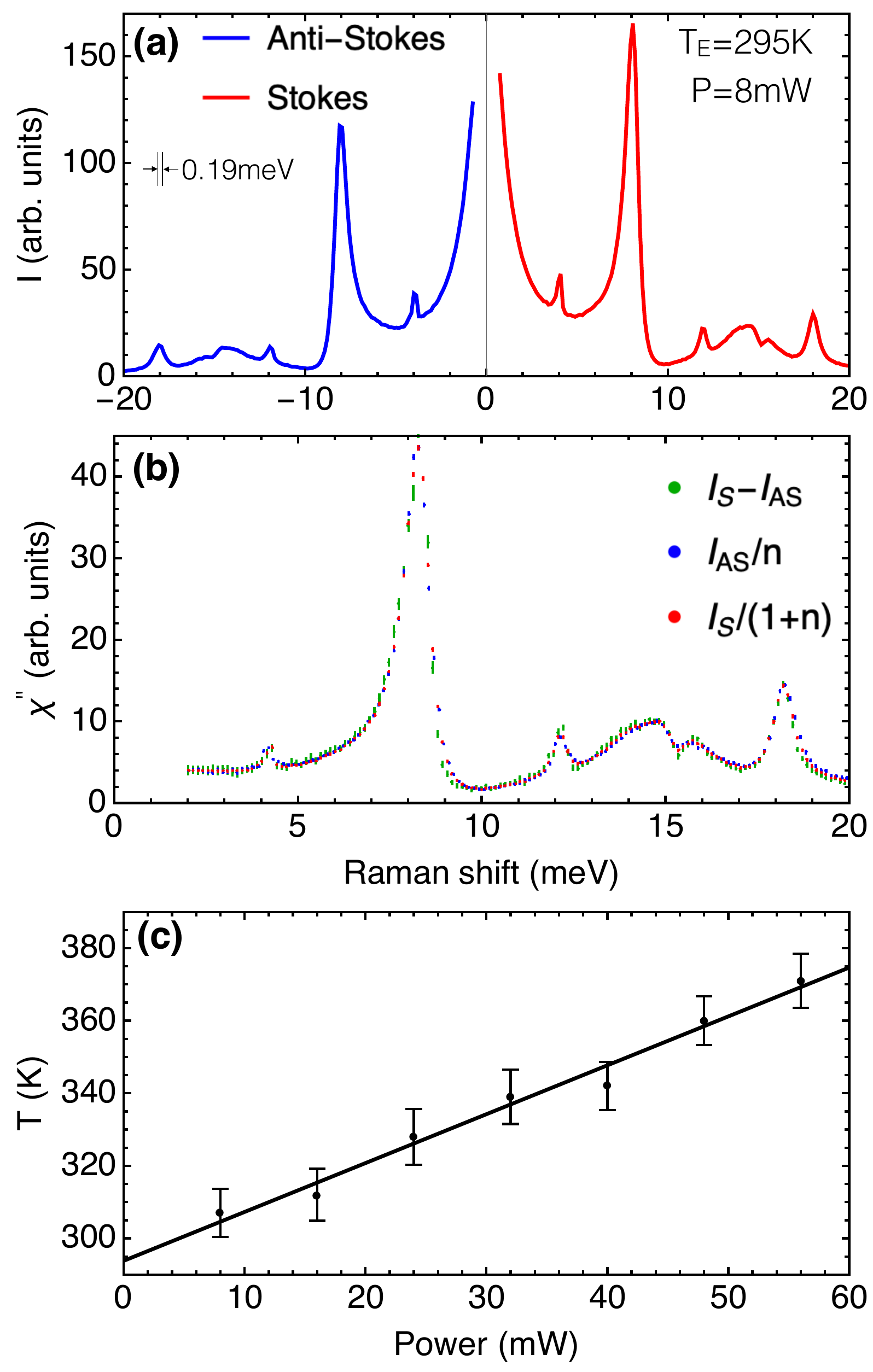}
	\centering
	\caption{\label{fig:SAS} 
		Stokes-Anti-Stokes analysis of the spectra taken in $ac$ polarization geometry~\cite{mai2021}:
		(a) Measured Raman cross-section $I(\omega)$ that includes both the Stokes (red line, $\omega>0$) and Anti-Stokes (blue line, $\omega<0$) parts measured at environmental temperature $T_{\mathrm{E}}$\,=\,295\,K with laser power P\,=\,8\,mW. (b) The Raman response $\chi''(\omega)$ calculated from the measured spectra in (a) [see Eq. \ref{eq:chidef}]. The laser-spot temperature, which appears in the Bose factor $n$, is 307\,K. The response obtained using three equivalent ways in Eq. \ref{eq:chidef} agree well, showing that the detailed balance relation Eq. \eqref{eq:SAS} holds. The spectral resolution is 0.19\,meV for panels (a-b). (c) The laser-power dependence of the laser-spot temperature (Eq. \ref{eq:T}). The straight line represents a linear fit corresponding to heating rate of 1.29$\pm$0.17\,K\,mW$^{-1}$.}
\end{figure}

The Raman susceptibility can then be obtained in three equivalent ways (using Eq. \ref{eq:SAS}):
\begin{equation}
\chi''(\omega) = \frac{I_{\mathrm{S}}(\omega)}{n(\omega)+1} = \frac{I_{\mathrm{AS}}(\omega)}{n(\omega)} = I_{\mathrm{S}}(\omega)-I_{\mathrm{AS}}(\omega)
\label{eq:chidef}
\end{equation}

We studied the Stokes and Anti-Stokes cross section relation for the $ac$ scattering geometry at 295\,K environmental temperature with varying laser power. 
In Fig.~\ref{fig:SAS} we show the results measured with 8\,mW laser power as an example. 
The Raman responses calculated from Stokes and Anti-Stokes cross sections match well with the temperature at the laser spot being 307\,K. 
We use Eq.~(\ref{eq:T}) with the phonon intensity integrated from 7.5 to 8.5\,meV to calculate the temperature at the laser spot. 
By a linear fit to the power dependence of the laser-spot temperature, we find the heating rate to be 1.29$\pm$0.17\,K\,mW$^{-1}$.

\subsection{Fitting model for the Raman susceptibility}
\label{app:fano}

\subsubsection{Temperatures above $T_{\mathrm{c}}$}

Here we describe the procedure we used to fit the $ac$ Raman susceptibility around $T_{\mathrm{c}}$ characterized by strongly asymmetric features, see Fig. 2d and Fig. 3 of the main text. In particular, it is well known that such asymmetry cannot be captured by a conventional Lorentzian oscillator lineshape and instead signifies an interference effect arising from interaction between a sharp mode and an excitation continuum \cite{fano1961,klein1983}. The general result following from Fermi's Golden Rule \cite{klein1983} is
\begin{equation}
\begin{gathered}
I(\omega) = {\rm Im} \sum_{i,j = {\mathrm{e,p}}} \left\{ t_i \langle i | G(\omega+i\delta) | j \rangle t_j\right\},
\\
G(z) = 
\begin{bmatrix}
G_{\mathrm{p}}^{-1}(z) & v\\
v & G_{\mathrm{e}}^{-1}(z)
\end{bmatrix}^{-1},
\end{gathered}
\label{eq:klein}
\end{equation}
where $t_{\mathrm{e/p}}$ are the Raman excitation matrix elements and $G_{\mathrm{e/p}i}(\omega)$ are the Green's functions of the electronic continuum and of the phonon mode (such that the corresponding susceptibilities of the individual components are given by $t_{\mathrm{e/p}i}^2G_{\mathrm{e/p}i}(\omega)$) and $v$ is the interaction matrix element between them (which is approximated by a frequency-independent constant). In the simplest case, for an infinite continuum with a constant density of states $\rho$ (such that $G_{\mathrm{e}}(\omega+i\delta) = i \pi \rho$) and a delta-function-like sharp mode $G_{\mathrm{p}}^{-1}(\omega+i\delta) = \omega_{\mathrm{p}}-\omega$ one gets \cite{blumberg1994,shangfei2020}:
\begin{equation}
I(\omega) = \frac{t_{\mathrm{e}}^2\pi \rho (\omega_{\mathrm{p}}-\omega-v t_{\mathrm{p}}/t_{\mathrm{e}})^2}{(\omega_{\mathrm{p}}-\omega)^2+(v^2\pi \rho)^2},
\end{equation}
where it is evident that the parameter $v t_{\mathrm{p}}/t_{\mathrm{e}}$ controls the asymmetry of the peak lineshape. Note that the total intensity can be smaller than the continuum intensity: e.g. at $\omega=\omega_{\mathrm{p}}-v t_{\mathrm{p}}/t_{\mathrm{e}}$ the total intensity vanishes ("Fano antiresonance"), while the continuum intensity at this frequency is equal to $t_{\mathrm{e}}^2\pi \rho$. This effect highlights the role of interference in the Fano model and a similar effect is observed in our data (Fig. 3c of the main text, around $10$ meV).

To describe the results obtained we need to adjust this simple model in the following ways. First, in our results we have observed three distinct phonon peaks. Taking into account their rather close spacing we need to generalize the above to include three sharp independent modes and take the interaction between continuum and each of the modes into account.

For the bare phonon Green's function, we use %~\cite{Mattuck1976}
\begin{equation}
G_{\mathrm{p}i}(\omega)=-\left(\frac{1}{\omega-\omega_{\mathrm{p}i}+i\gamma_{\mathrm{p}i}}-\frac{1}{\omega+\omega_{\mathrm{p}i}+i\gamma_{\mathrm{p}i}}\right)~,
\label{eq:chiphon}
\end{equation}
such that ${\rm Im} \chi_{\mathrm{p}i}(\omega)$ is an odd function of $\omega$ (the requirement $\chi''(-\omega) = -\chi''(\omega)$ arises due to the requirement of the real-time response to be real). The parameters $\omega_{\mathrm{p}i}$ and $\gamma_{\mathrm{p}i}$ respectively have the physical meaning of the mode energy and the half width at half maximum (HWHM).

Finally, the electronic continuum contribution has been found to be best described by a purely relaxational response $\chi_{\mathrm{cont}}^{-1}(\omega) \propto -i \Gamma \omega +\omega_0^2$ that correspond to the overdamped ($\Gamma\gg\omega_0$) limit of the Drude-Lorentz model. This form follows from the time-dependent Ginzburg-Landau \cite{goldenfeld2018} description of the dynamics of the electronic order parameter $\varphi$, which can be applied to a diverse range of systems from charge-density wave \cite{yusupov2010,zong2019} to superconducting \cite{Schuller2006} ones. In our case, due to absence of a gap above $T_{\mathrm{c}}$, the dynamics of the order parameter can be expected to be of relaxational type, resulting in the following time-dependent Ginzburg-Landau equation for the electronic order parameter $\varphi$:
\begin{equation}
\begin{gathered}
-\Gamma \partial_t \varphi = \frac{d F}{d\varphi},
\\
F[\varphi,T] =  a(T) \frac{\varphi^2}{2}+ b\frac{\varphi^4}{4},
\end{gathered}
\end{equation}
where $a(T)>0$ for $T>T_{\mathrm{c}}^{\mathrm{el}}$ and $a(T_{\mathrm{c}}^{\mathrm{el}})=0$. Above $T_{\mathrm{c}}$ the quartic term in $F[\varphi,T]$ can be neglected resulting in $-\Gamma \partial_t \varphi  = a(T) \varphi$. For an external source (which is the quadrupolar field of $xz$ symmetry provided by the photons in the Raman process), the response function becomes:
\[
\chi_{\mathrm{el}}(\omega, T) = \frac{C}{i\Gamma \omega+a(T)}  = \frac{C/\Gamma}{i\omega+a(T)/\Gamma} ,
\]
which is determined by two parameters: $C/\Gamma$ and $a(T)/\Gamma$. As the above form resembles an overdamped oscillator, we have used the notation $a(T)\equiv \omega_0^2(T)$ for illustrative purposes. Furthermore, for the case of a conventional exciton, the response is expected to be that of a usual oscillator, similar to phononic modes; we found this notation useful to explain the physics of this mode. Thus, we have used the
following form of the continuum response:
\begin{equation}
G_{\mathrm{e}}^{-1}(\omega) = -i\omega+\Omega_{\mathrm{e}},
\label{eq:chiel}
\end{equation}
where $\Omega_{\mathrm{e}} =\frac{\omega_0^2}{\Gamma}$.	

Summarizing the above, we use the following model to fit the data:
\begin{equation}
\chi^{\prime\prime}(\omega) = {\rm Im} \sum_{ij} T_iG_{ij}(\omega)T_j~,
\label{eq:fanoChi}
\end{equation}
where $T=\left(\begin{array}{cccc}t_{p1}&t_{p2}&t_{p3}&t_{e}\end{array}\right)$ is the matrix element of the Raman light scattering process, and $G$ is defined by
\begin{equation}
G=(G_0^{-1}-V)^{-1}.
\label{eq:fanoG}
\end{equation}
where
\begin{equation}
G_0=\begin{pmatrix}G_{p1}&0&0&0\\0&G_{p2}&0&0\\0&0&G_{p3}&0\\0&0&0&G_{e}\end{pmatrix}~,
\label{eq:fanoG0}
\end{equation}
\begin{equation}
V=\begin{pmatrix}0&0&0&v_1\\0&0&0&v_2\\0&0&0&v_3\\v_1&v_2&v_3&0\end{pmatrix}~.
\label{eq:fanoV}
\end{equation}
This model can be thought of as a generalization of \eqref{eq:klein} for three phonons, where $G_{\mathrm{p}i}^{-1}$ (i=1,2,3) correspond to the B$_{2g}^{(i)}$ phonon mode and have the form \eqref{eq:chiphon}, and the parameter $v_i$ describe the coupling strength of this mode to the electronic excitations which are represented by $G_{e}^{-1}$ that has the form \eqref{eq:chiel}.

\begin{figure}[h!]
	\centering
	\includegraphics[width=0.5\textwidth]{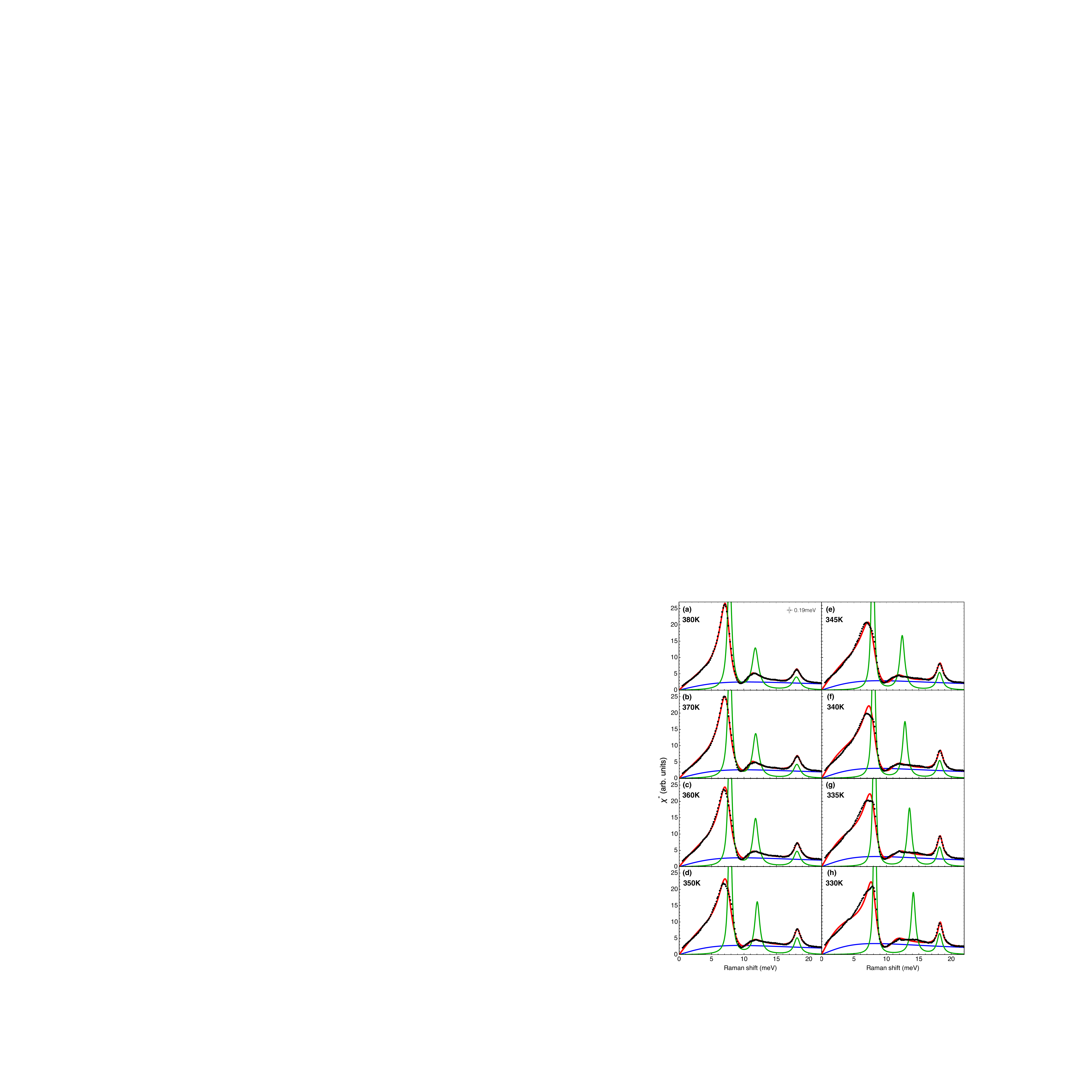}
	\caption{Temperature dependence of Raman response $\chi^{\prime\prime}$ in the $ac$ scattering geometry for Ta$_2$NiSe$_5$~\cite{mai2021}. The data is shown by black dots, along with the Fano fits [Eq.~(\ref{eq:fanoChi})] shown as red curves. The spectral resolution is 0.19\,meV. The bare optical phonon mode response is shown by green curves, and the bare excitonic continuum is shown by blue curves [see Eq.~(\ref{eq:chibare})].}
	\label{fig:Linear}
\end{figure}

In Fig. \ref{fig:Linear} (red lines) we present the obtained fits using Eq. (\ref{eq:fanoG}-\ref{eq:fanoV}) (for parameters see Sec. 1.3.3) for temperatures above $T_{\mathrm{c}}$ in linear scale (as opposed to logarithmic used in Fig. 3 of the main text). Furthermore, we define the bare excitonic (blue line) and phononic (green line) responses as:
\begin{equation}
\chi_{\mathrm{\mathrm{opt}}}''(\omega) = \sum_{i=1}^3 t_{\mathrm{p}i}^2 {\rm Im} G_{\mathrm{p}i}(\omega);
\;
\chi_{\mathrm{cont}}''(\omega)  = \frac{t_{\mathrm{e}}^2 \omega}{\omega^2 + \Omega_{\mathrm{e}}^2}.
\label{eq:chibare}
\end{equation}
One observes that the fit quality away from $T_{\mathrm{c}}$ is excellent, but becomes slightly worse on approaching $T_{\mathrm{c}}$. This can be related to the presence of nonlinear exciton-phonon interactions that are not included in the Fano model and may become important close to the transition (see also discussion on top of page 7 of the main text).

\subsubsection{Temperatures below $T_{\mathrm{c}}$}
At $T<T_{\mathrm{c}}$ two important qualitative changes in the spectra occur. First, due to the symmetry breaking the selection rules change (see Methods), such that the modes of $A_{\mathrm{g}}$ symmetry can appear in $ac$ polarization geometry. For the $ac$ spectra below 22 meV (where the Fano model fit is made), four additional modes appear - corresponding to A$_{g}^{(1)}$-A$_{g}^{(4)}$ modes. We perform three steps to account for these additional features: First, we use the phononic lineshape Eq. \eqref{eq:chiphon} to fit the corresponding features in $aa$ polarization geometry to obtain the frequencies and HWHM for each of the A$_{g}^{(1)}$-A$_{g}^{(4)}$ modes [Fig. \ref{fig:Ag} (a-d)]. Second, we use Eq. \eqref{eq:chiphon} (see also Eq. (1) in \cite{mai2021}) to individually fit the three "leakage" modes in $ac$ polarization geometry, with the frequencies and HWHM fixed; namely, the only free parameter is the light-scattering vertex. The vertex values obtained are shown in Fig. \ref{fig:Ag} (e-h). Third, we subtract the associated response from the total one and then fit the remaining response with the model discussed below.

\begin{figure}[h!]
	\centering
	\includegraphics[width=0.5\textwidth]{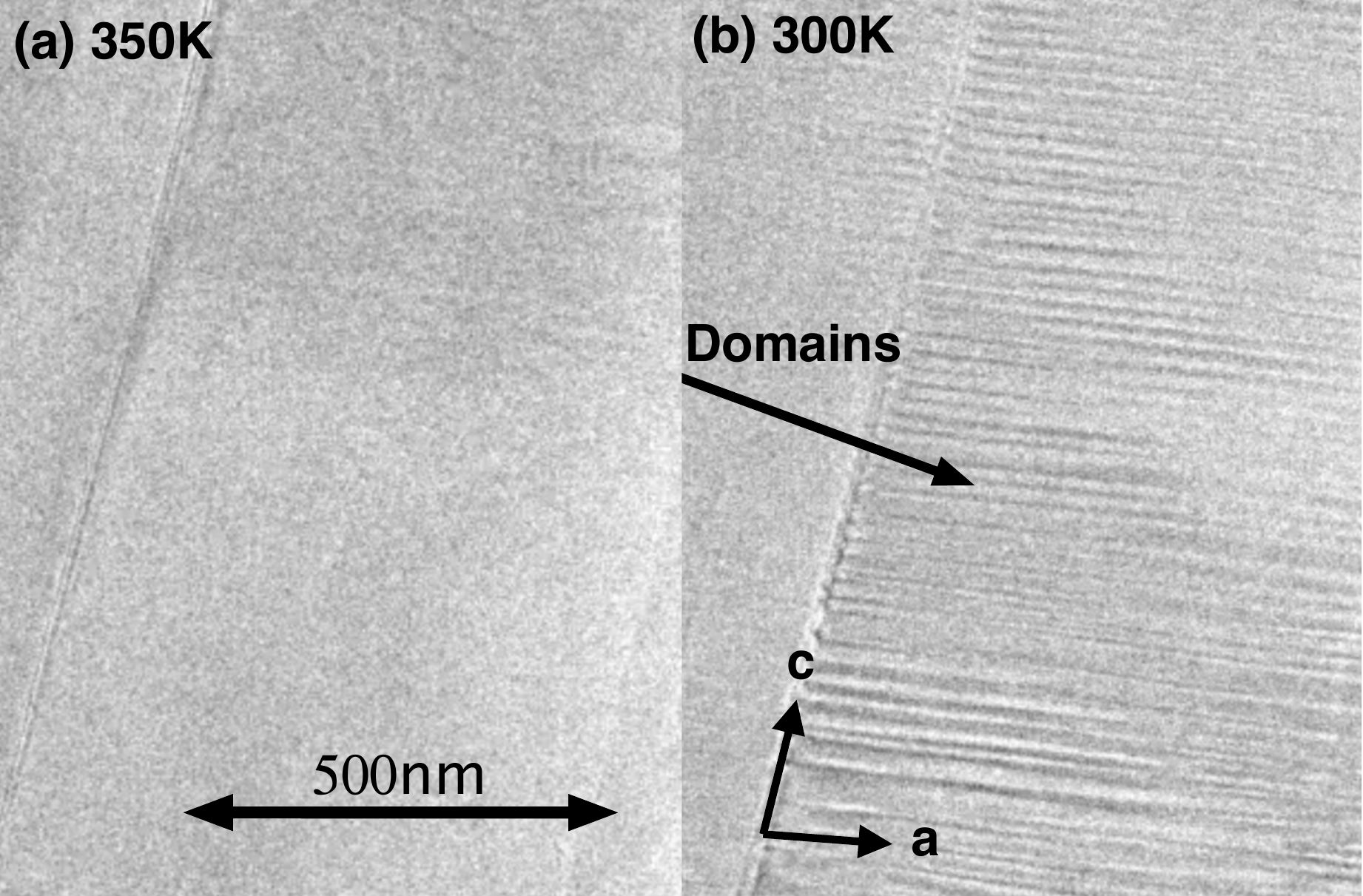}
	\caption{Transmission electron microscopy images of the sample~\cite{mai2021}: (a) at $350$ K and (b) at $300$ K. Structural domains with average spacing on the order of sizes of 200\,$\AA$\, can be directly observed for temperature below $T_{\mathrm{c}}$.}
	\label{fig:domains}
\end{figure}

Additionally, not too far away from $T_{\mathrm{c}}$, we have found that an additional enhancement of the low-energy response. We relate this observation with a quasiperiodic domain structure formed below $T_{\mathrm{c}}$. In Fig. \ref{fig:domains} we present transmission electron microscopy images of the sample. Thin domains along the $a$ axis are resolved below $T_{\mathrm{c}}$, but not above $T_{\mathrm{c}}$. The average spacing $\bar{d}$ between these stripes along $c$-axis direction is on the order of 200\,$\AA$. Acoustic phonon, scattering on this quasiperiodic structure, can take a recoil momentum of the order $2\pi/\bar{d}$; this way a finite $q_d=\pm2\pi/\bar{d}$ phonon can contribute to the Raman intensity (where the momentum transfer is approximately zero). At finite $q$, introducing the acoustic phonon coordinate $x_q = \sqrt{q} u_q$, where $u_q$ - is the Fourier transform of the atomic displacement amplitude, one obtains that the Green's function of the acoustic mode:
\begin{equation}
G_{s}(\omega, q)=-(\frac{1}{\omega-c_{\mathrm{s}} q+ir_{\mathrm{s}} q}-\frac{1}{\omega+c_{\mathrm{s}} q+ir_{\mathrm{s}} q})~.
\label{eq:fanoS}
\end{equation}
Its light-scattering vertex $t_{\mathrm{s}}(q)$ and coupling to the excitonic continuum $v_{\mathrm{s}}(q)$ are both proportional to the square root of wavevector: $t_{\mathrm{s}} = \tau_{\mathrm{s}} \sqrt{q}$ and $v_{\mathrm{s}} = \beta_{\mathrm{s}} \sqrt{q}$. The frequency $\omega_{\mathrm{s}}$ and the HWHM $\gamma_{\mathrm{s}}$ are both proportional to the wavevector: $\omega_{\mathrm{s}} = c_{\mathrm{s}} q$ and $\gamma_{\mathrm{s}} = r_{\mathrm{s}} q$, where $c_{\mathrm{s}}$ is the sound velocity, that can be deduced from the X-ray measurements \cite{XRay2018} to be $c_{\mathrm{s}} \approx 30$\,meV\,\AA\, away from $T_{\mathrm{c}}$. After introducing this mode, Eqs.~(\ref{eq:fanoG0}-\ref{eq:fanoV}) extend to
\begin{equation}
G_0'=\begin{pmatrix}G_{p1}^0&0&0&0&0\\
0&G_{p2}^0&0&0&0\\
0&0&G_{p3}^0&0&0\\
0&0&0&G_{s}&0\\0&0&0&0&G_{e}\end{pmatrix}\,,
\label{eq:fanoG0LT}
\end{equation}
\begin{equation}
V'=\begin{pmatrix}
0&0&0&0&v_1\\
0&0&0&0&v_2\\
0&0&0&0&v_3\\
0&0&0&0& \beta_{\mathrm{s}} \sqrt{q}\\
v_1&v_2&v_3 & {\beta_{\mathrm{s}}} \sqrt{q} &0\end{pmatrix}\,,
\label{eq:fanoVLT}
\end{equation}
and the vertex for light-scattering process becomes $T'=\left(t_{p1},t_{p2},t_{p3},\tau_{\mathrm{s}} \sqrt{q},t_{e}\right)$.

Then the total response function containing coupling to acoustic mode for domain walls $i$ lattice constants apart is given by
\begin{equation}
\chi^{\prime\prime}_i\sim\Im T^{'T}G'T'~.
\label{eq:fanoChii}
\end{equation}

However, as has been noted above, the domain size is not the same for all domains; to take this into account we will assume that the spacing between stripes $i$ in units of the crystal unit cell constant $c$ follows the Poisson distribution
\begin{equation} 
P(i; \mu = \frac{\bar{d}}{c}) = \frac{\mu^i e^{-\mu}}{i!}, 
\label{eq:Poisson}
\end{equation}
where $\mu$ is average inter-domain distance in the number of $c$-direction unit cells.  Recognizing that the low-frequency feature is inhomogeneously broadened by random distribution of the stripe distances, for the fitting procedure to the measured response function we perform summation over the distribution function:
\begin{equation} 
\chi^{\prime\prime}(\omega)=\sum_{i=1}^{\infty}P(i; \mu)\chi^{\prime\prime}_i(\omega),
\label{eq:ChiAcous}
\end{equation}
in which each $\chi^{\prime\prime}_i(\omega)$ is the full response function containing coupling to acoustic mode for domain walls $i$ lattice constants apart. For fitting, we have taken $\beta = 7.7$\,meV\,\AA$^{\frac{1}{2}}$ from the assumption that the static susceptibility including coupling to both acoustic and optical phonons should diverge at $T_{\mathrm{c}}$ (see Section S 1.3.4).

\subsubsection{Fano fit parameters}

The fits to the data, as well as the phononic and excitonic components, are shown in Fig. 3 of the main text.  
For the 300\,K and 265\,K spectra below $T_{\mathrm{c}}$ we also include the acoustic contribution due to coupling via quasi-periodic structure of domain walls. The value $\tau_{\mathrm{s}}$ is 6.0\,meV\,\AA$^{\frac{1}{2}}$ at 300\,K and 2.2\,meV\,\AA$^{\frac{1}{2}}$ at 265\,K; the value $r_{\mathrm{s}}$ is 45\,meV\,\AA\, at both 300\,K and 265\,K.

\begin{figure}[h!]
	\centering
	\includegraphics[width=0.46\textwidth]{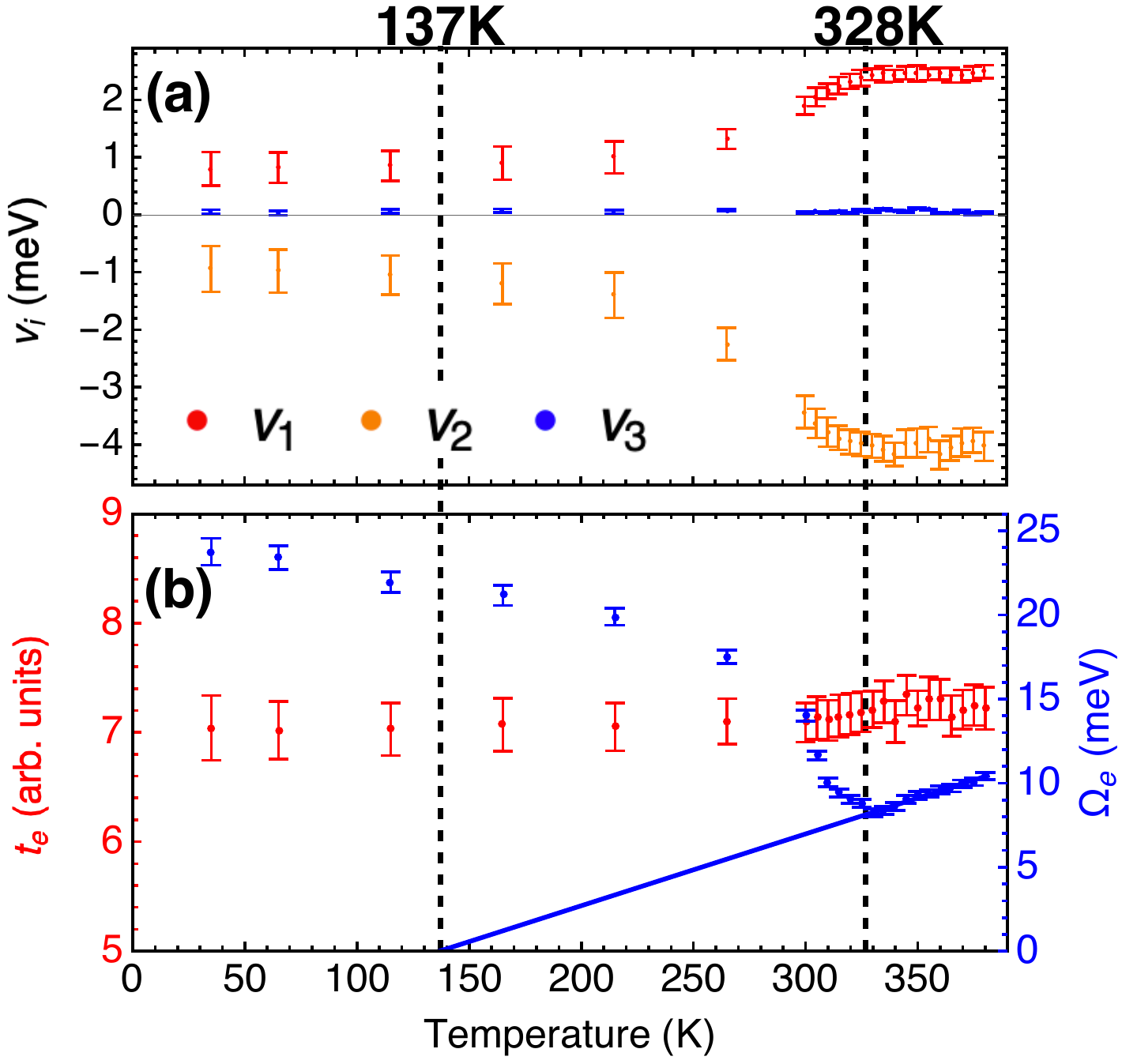}
	\caption{Temperature dependence of (a) the coupling strengths $v_{1,2,3}$ of the three $B_{\mathrm{2g}}$ optical phonons to the excitonic continuum, and (b) the light-scattering vertex $t_{\mathrm{e}}$ and the parameter $\Omega_{\mathrm{e}}$ of the excitonic continuum~\cite{mai2021}. These values are obtained from the Fano lineshape fit, Eqs. (\ref{eq:fanoChi}-\ref{eq:fanoV},\ref{eq:ChiAcous}).}
	\label{fig:S1}
\end{figure}

\begin{figure}[h!]
	\centering
	\includegraphics[width=0.5\textwidth]{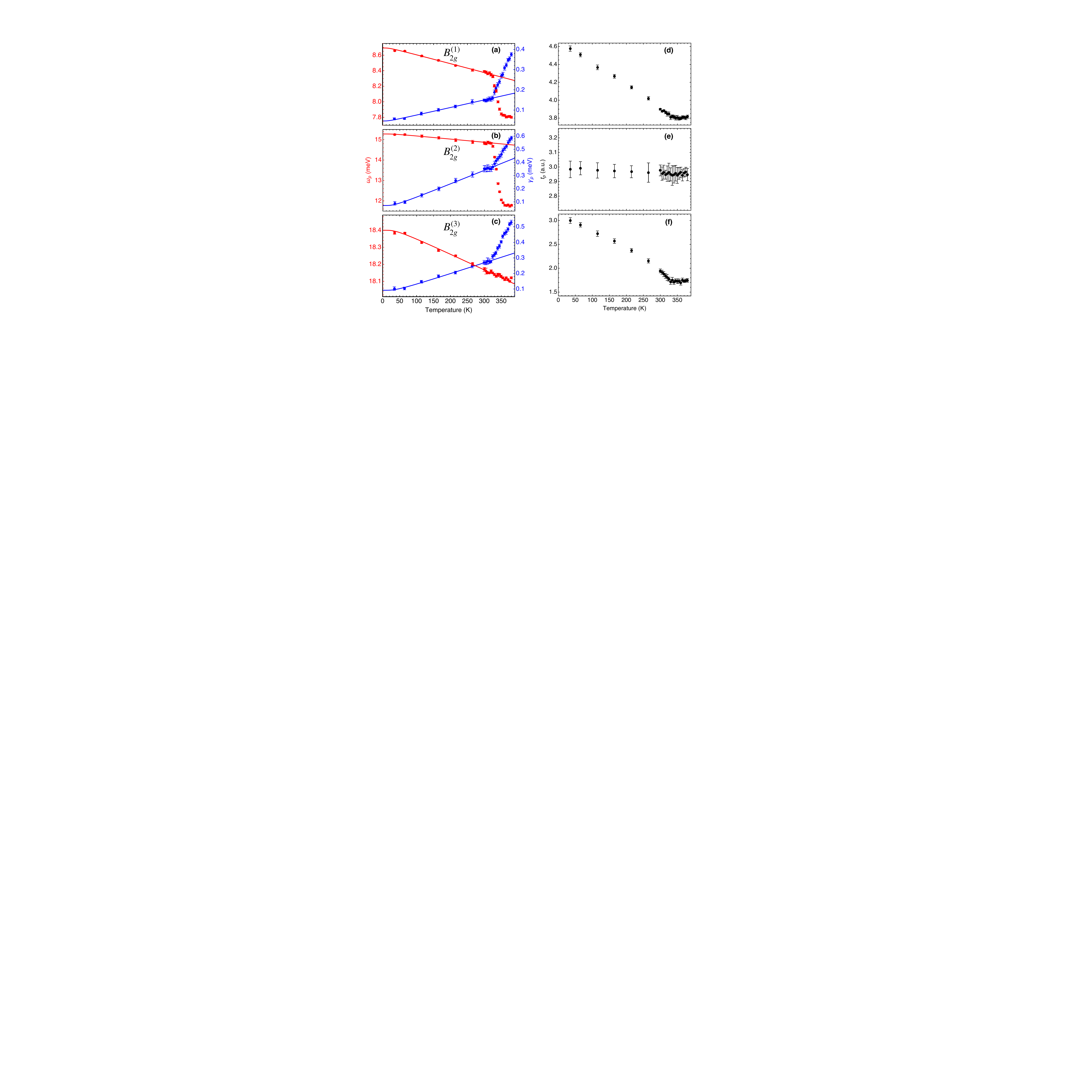}
	\caption{Temperature dependence of (a)-(c) bare frequencies ($\omega_{\mathrm{p}}$) and HWHM ($\gamma_{\mathrm{p}}$), and (d)-(f) light-scattering vertex ($t_{\mathrm{p}}$) of the three $B_{\mathrm{2g}}$ optical phonons obtained from the Fano lineshape fit Eq. \eqref{eq:fanoChi}~\cite{mai2021}.}
	\label{fig:S2}
\end{figure}

The temperature dependence of the coupling strengths $v_i$, the light-scattering vertex $t_{\mathrm{e}}$ of the excitonic continuum, and the parameter $\Omega_{\mathrm{e}}$ of the continuum is shown in Fig. \ref{fig:S1}. The frequency, linewidth, and light-scattering vertex of the three $B_{\mathrm{2g}}$ phonon modes as a function of temperature are given in Fig. \ref{fig:S2}. In Fig. \ref{fig:S1} (a), one observes that the continuum-phonon coupling strengths $v_{1,2}$ are almost constant above $T_{\mathrm{c}}$ and decrease in magnitude at lower temperatures. The third phonon is almost decoupled from the continuum at all temperatures, and essentially have a symmetric lineshape (see Fig. 3 of the main text). The signs of $v_{1,2}$ reflect the different shapes of the resulting asymmetric features: the $B_{\mathrm{2g}}^{(1)}$ phonon peak is skewed to the left, while $B_{\mathrm{2g}}^{(2)}$ is skewed in the opposite direction (see Fig. 2d and Fig. 3 of the main text). In Fig. \ref{fig:S1} (b), one can see that the light-scattering vertex of the excitonic continuum is temperature independent. The parameter $\Omega_{\mathrm{e}}$ has the same temperature dependence as the inverse bare static excitonic susceptibility (see Fig. 4 of the main text).

In Fig. \ref{fig:S2} (a) - (c) we show the frequency $\omega_{\mathrm{p}}$ and HWHM $\gamma_{\mathrm{p}}$ of the three $B_{\mathrm{2g}}$ phonon modes. The frequencies increase on cooling, with the lower-energy $B_{\mathrm{2g}}^{(1,2)}$ modes showing a pronounced increase around $T_{\mathrm{c}}$. In the same region, all the three modes show a pronounced decrease of HWHM on cooling, suggesting an enhanced phonon scattering above $T_{\mathrm{c}}$. Below 300\,K, the temperature dependence of both frequency and HWHM of the phonon modes is in accordance with a simple model assuming anharmonic decay into two phonons with identical frequencies and opposite momenta~\cite{Klemens1966}:

\begin{equation}
\omega(T)=\omega_0-\omega_2[1+\frac{2}{e^{\hbar\omega_0/2k_{\mathrm{B}}T}-1}],
\label{eq:energyTwo}
\end{equation}
and
\begin{equation}
\gamma(T)=\gamma_0+\gamma_2[1+\frac{2}{e^{\hbar\omega_0/2k_{\mathrm{B}}T}-1}].
\label{eq:gammaTwo}
\end{equation}

The light-scattering vertex ($t_{\mathrm{p}}$) as a function temperature is presented in Fig. \ref{fig:S2} (d)-(f). Below $T_{\mathrm{c}}$, the vertex values increase on cooling, most significantly for the $B_{\mathrm{2g}}^{(3)}$ mode, reflecting the changes in the structure of these modes and the corresponding atomic displacement patterns below $T_{\mathrm{c}}$.

\begin{figure}[h!]
	\centering
	\includegraphics[width=0.4\textwidth]{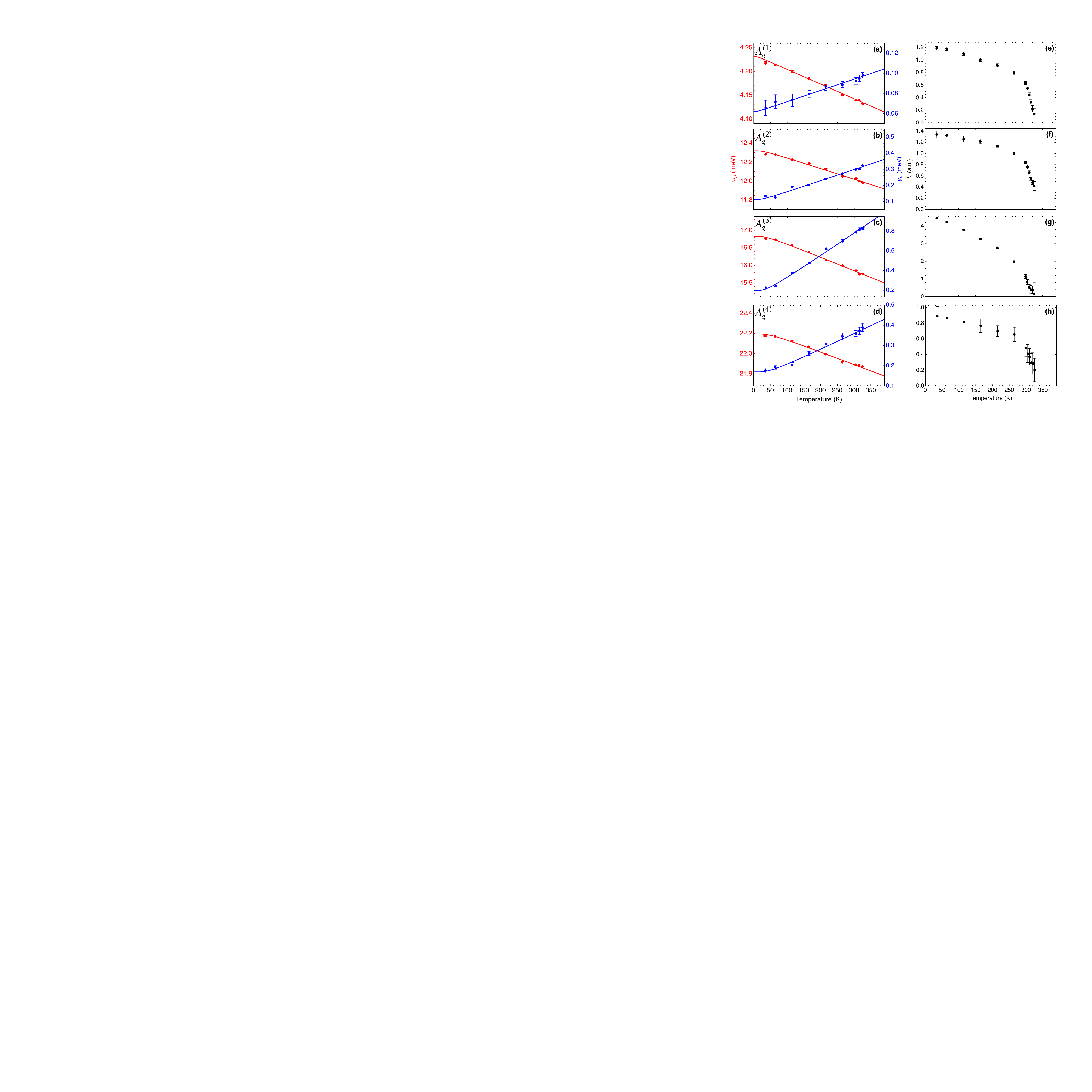}
	\caption{Temperature dependence of (a)-(d) bare frequencies ($\omega_{\mathrm{p}}$) and HWHM ($\gamma_{\mathrm{p}}$), and (e)-(h) light-scattering vertex ($t_{\mathrm{p}}$) of the four $A_{\mathrm{g}}$ optical phonons obtained from the phononic lineshape fit Eq. \eqref{eq:chiphon}~\cite{mai2021}.}
	\label{fig:Ag}
\end{figure}

In Fig. \ref{fig:Ag} we present the parameters used to subtract the additional A$_{g}^{(1)}$-A$_{g}^{(4)}$ modes appearing in the $ac$ spectra at $T<T_{\mathrm{c}}$. The temperature dependence of the frequency and HWHM is consistent with the anharmonic decay model; the vertex values become non-zero below the transition temperature, and tend to saturate at very low temperature.

\subsubsection{Static susceptibility}
The function (\ref{eq:fanoChi},\ref{eq:fanoG}) used to fit the finite-energy Raman response (by taking the imaginary part) can be also used to calculate the static $B_{\mathrm{2g}}$ susceptibility by simply taking its zero-frequency limit:
\begin{equation}
\chi_{\mathrm{ac}}(0,T) =\sum_{ij} T_iG_{ij}(0)T_j.
\label{eq:chitot}
\end{equation}
This expression is used to calculate the combined susceptibility data for Fig. 5 in the main text.
Additionally, one can define the purely electronic (continuum) and purely phononic parts of the susceptibility as:
\begin{equation}
\begin{gathered}
\chi_{\mathrm{cont}}(0,T) =\frac{t_{\mathrm{e}}^2(T)}{\Omega_{\mathrm{e}}(T)},
\\
\chi_{\mathrm{opt}}(0,T) =\sum_{i=1}^3\frac{t_{\mathrm{p}i}^2(T)}{2\omega_{\mathrm{p}i}(T)}.
\end{gathered}
\end{equation}
The critical temperature due to excitons and optical phonons $T^{comb}_{\mathrm{c}}$ (excluding the coupling to strain modes) can be obtained from Eq. \eqref{eq:chitot}; the condition for $\chi_{\mathrm{ac}}(0,T)$ to diverge is independent of the light-scattering vertex $T_i$, and is given by $\det G_{ij}^{-1}(0) = 0$ resulting in the following equation:
\begin{equation}
\Omega_{\mathrm{e}}(T^{comb}_{\mathrm{c}}) - \sum_{i=1}^3\frac{2 v_i^2}{\omega_{\mathrm{p}i}} = 0.
\end{equation}
Using the values of $v_i$ and $\omega_{\mathrm{p}i}$ above $350$ K where they do not depend strongly on temperature and a linear extrapolation of $\Omega_{\mathrm{e}}(T)$ (Fig. \ref{fig:S1}) one obtains $T^{comb}_{\mathrm{c}}\approx 227$ K, consistent with Fig. 5b, where a linear extrapolation for $\omega_{\mathrm{p}i}$ (Fig. 4b of the main text, dashed lines) has been additionally used.

One can also include the contribution of acoustic phonons at ${\bf q}\to0$ to get the full susceptibility. Performing the same calculations as above, but using Eq.\,\ref{eq:fanoG0LT} for the Green's function and \ref{eq:fanoVLT} for the coupling matrix with infinitesimal $q$ one gets:
\begin{equation}
\Omega_{\mathrm{e}}(T_{\mathrm{c}}) - \sum_{i=1}^3\frac{2 v_i^2}{\omega_{\mathrm{p}i}}  - \frac{2\beta_{\mathrm{s}}^2}{c_{\mathrm{s}}}= 0.
\end{equation}
One observes that the effect of the acoustic mode on $T_{\mathrm{c}}$ does not vanish in the $q\to0$ limit despite the coupling is $\sim \sqrt{q}$. This equation allows to determine $\beta_{\mathrm{s}}$ used for fitting the domain-induced low-energy upturn below $T_{\mathrm{c}}$ discussed above.

The combined (excitonic interacting with phononic, red symbols in Fig.\,5 of main text) static susceptibility could be also obtained by performing Kramers-Kronig transformation directly on the experimentally measured data. 
We have restricted the integration to 100 meV at high energy. 
A linear extrapolation of the data to zero intensity below 0.4 meV cutoff was used. 
In Table\,\ref{table:chi} we compare the static susceptibility calculated from the analytical Fano model to that obtained by Kramers-Kronig numerical transformation directly from the data. 
We note that the values obtained by numerical integration are systematically smaller than the values obtained from the analytical Fano model because of finite cutoff frequency (100\,meV). 
The static susceptibility values derived by two methods are in good agreement.

\begin{table}
	\begin{center}
		\caption{
			\label{table:chi}
			The combined static $B_{\mathrm{2g}}$ susceptibility at three temperatures, calculated by two methods: using the analytical expression from the Fano model with the fitting parameters or by numerical Kramers-Kronig transformation.}
		\vspace{2mm}
		\begin{tabular}{c|ccc}
			Temperature (K) & 380     & 330 & 310 \\
			\hline
			$\chi_{\mathrm{ac}}(0,T)$(a.u.) [Fano model]  & 15.0\;(0.7) & 19.6\;(1.2) & 18.4\;(0.9) \\
			$\chi_{\mathrm{ac}}(0,T)$(a.u.) [Numerical KK]   & 14.1 & 18.5 & 17.4 \\
		\end{tabular}
	\end{center}
\end{table}

\subsubsection{Softening of the $B_{\mathrm{2g}}$ acoustic mode above $T_{\mathrm{c}}$}
\label{app:acous}
Here we discuss the interpretation of the results of Nakano et al., \cite{XRay2018} in light of our data. In particular, we show that the softening of the acoustic mode observed in those experiments can be actually attributed to the coupling of the acoustic mode to the softening excitonic continuum. The renormalized modes can be obtained by solving the equation $\det\{G(\omega)^{-1}-V\} = 0$. As the relevant frequencies for the acoustic mode are much lower than that of the optical or excitonic modes, we can keep $\omega$ only in the bare acoustic mode's Green's function resulting in:
\begin{equation}
\omega^2 = \omega_{\mathrm{s}}^2-\frac{2v_{\mathrm{s}}^2\omega_{\mathrm{s}}}{\tilde{\Omega}_{\mathrm{e}}(T)},
\end{equation}
where
\begin{equation}
\tilde{\Omega}_{\mathrm{e}}(T) = \Omega_{\mathrm{e}}(T)-\sum_{i=1}^3\frac{2 v_i^2}{\omega_i}.
\end{equation}
Note that at $T_{\mathrm{c}}$ the total static susceptibility must diverge, resulting in a condition $\det\{G(\omega,T=T_{\mathrm{c}})^{-1}-V\} = 0$. This gives a constraint on the value of $v_{\mathrm{s}}$: $\Omega_{\mathrm{e}}(T_{\mathrm{c}})-\sum_{i=1}^3\frac{2 v_i^2}{\omega_i} = \frac{2v_{\mathrm{s}}^2}{\omega_{\mathrm{s}}}$. Applying this result to the equation above we get
\begin{equation}
\omega^2 = \omega_{\mathrm{s}}^2\left(1-\frac{\tilde{\Omega}_{\mathrm{e}}(T_{\mathrm{c}})}{\tilde{\Omega}_{\mathrm{e}}(T)}\right).
\end{equation}
Finally, substituting $\omega_{\mathrm{s}}=c_{\mathrm{s}}^0q$ one finds that the dispersion of the acoustic mode is given by $\omega = \tilde{c}_{\mathrm{s}}q$, where
\begin{equation}
\tilde{c}_{\mathrm{s}}(T) = c_{\mathrm{s}}^0\sqrt{1-\frac{\tilde{\Omega}_{\mathrm{e}}(T_{\mathrm{c}})}{\tilde{\Omega}_{\mathrm{e}}(T)}}.
\label{eq:csrenorm}
\end{equation}
Note that from the above it follows that at $T_{\mathrm{c}}$ the sound velocity softens to zero, as is seen experimentally \cite{XRay2018}, even in the absence of an intrinsic lattice instability, i.e. purely due to exciton-phonon coupling.

One can also estimate the expected renormalization of the sound velocity away from $T_{\mathrm{c}}$. Using the parameters extracted from the fits, we obtain $\tilde{c}_{\mathrm{s}}(T=400K)/c_{\mathrm{s}}^0\approx 0.65$. Experimentally, taking $c_{\mathrm{s}}^0$ to be the low-temperature value of $c_{\mathrm{s}}$ (which also agrees with DFT calculations) \cite{XRay2018} one obtains $\tilde{c}_{\mathrm{s}}(T=400K)/c_{\mathrm{s}}^0\approx 0.65$, in excellent agreement with the estimate from Eq. \eqref{eq:csrenorm}. Thus, our observations suggest that Ta$_2$NiSe$_5$ does not have a ferroelastic instability, and that the softening of the acoustic modes can be well described as being due to the coupling to the softening excitonic continuum instead.

\subsection{Additional details for low-energy Raman spectra}

\begin{figure}
	\centering
	\includegraphics[width=0.5\textwidth]{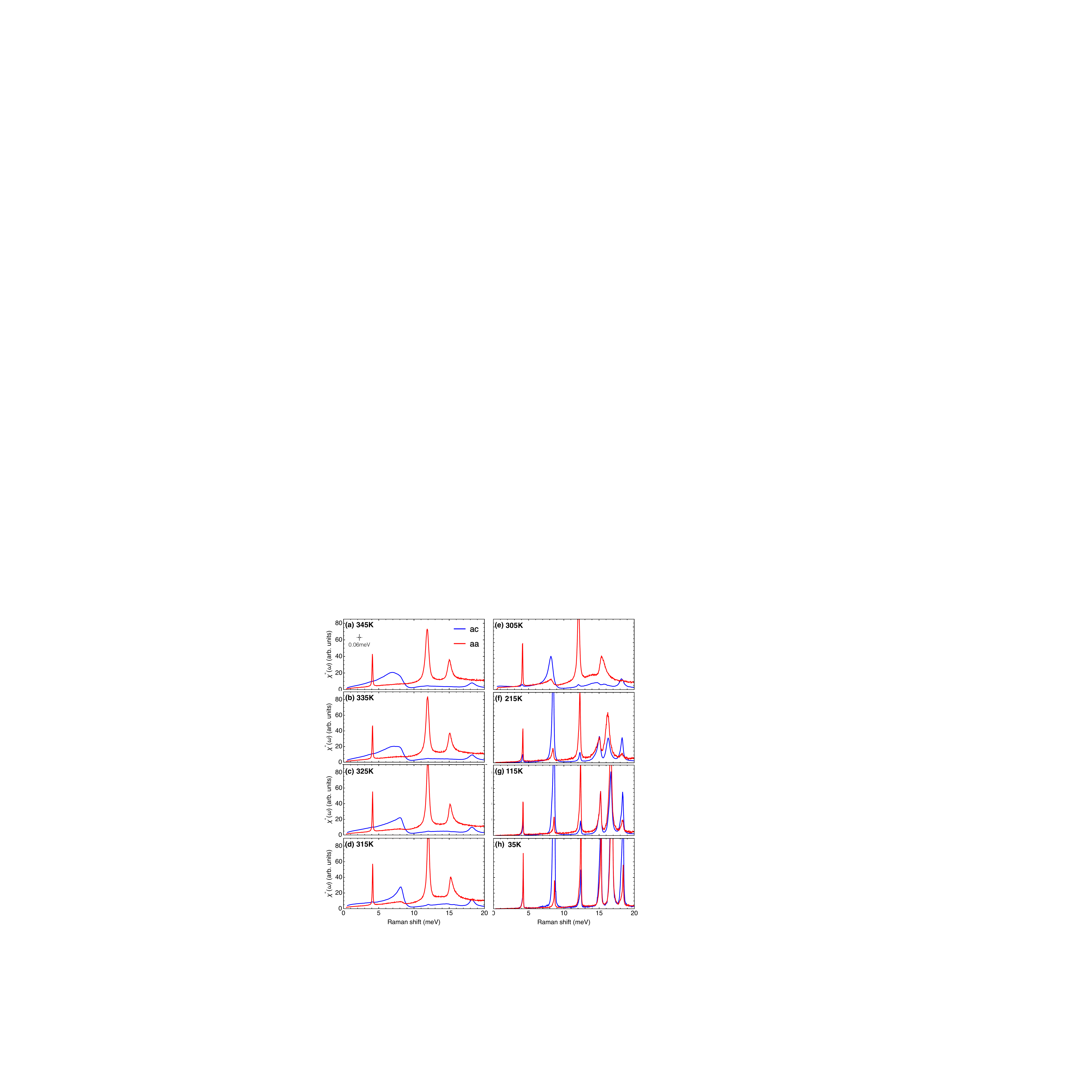}
	\caption{Comparison of the temperature dependencies of the low-frequency Raman response in $aa$ and $ac$ scattering geometries at temperatures between 345 and 35\,K.}
	\label{fig:aa}
\end{figure}

In Fig.\,2f of the main text we present the temperature dependence of the low-energy Raman response in $ac$ geometry. 
In Fig.\,\ref{fig:aa} we present the comparison of the low-energy Raman response in $aa$ and $ac$ geometries. One notes the absence of the strong Fano features in $aa$ geometry, comparable to those in $ac$ geometry (see also Fig.\,2f of the main text) at all temperatures. 
We also note that the $A_{\mathrm{g}}$ phonon modes exhibit no softening upon cooling towards the transition temperature; instead, their frequencies follow the anharmonic model between 360 and 35\,K~\cite{mai2021}. 
Additionally, the appearance of "leakage" between $aa$ and $ac$ geometries is clearly visible below $T_{\mathrm{c}}$ (Fig.\,\ref{fig:aa} d-h).

\subsection{Additional details for high-energy Raman spectra}
\label{app:highen}

\subsubsection{Details of the temperature dependence of the high-energy $ac$ spectra}

\begin{figure}[h!]
	\centering
	\includegraphics[width=0.5\textwidth]{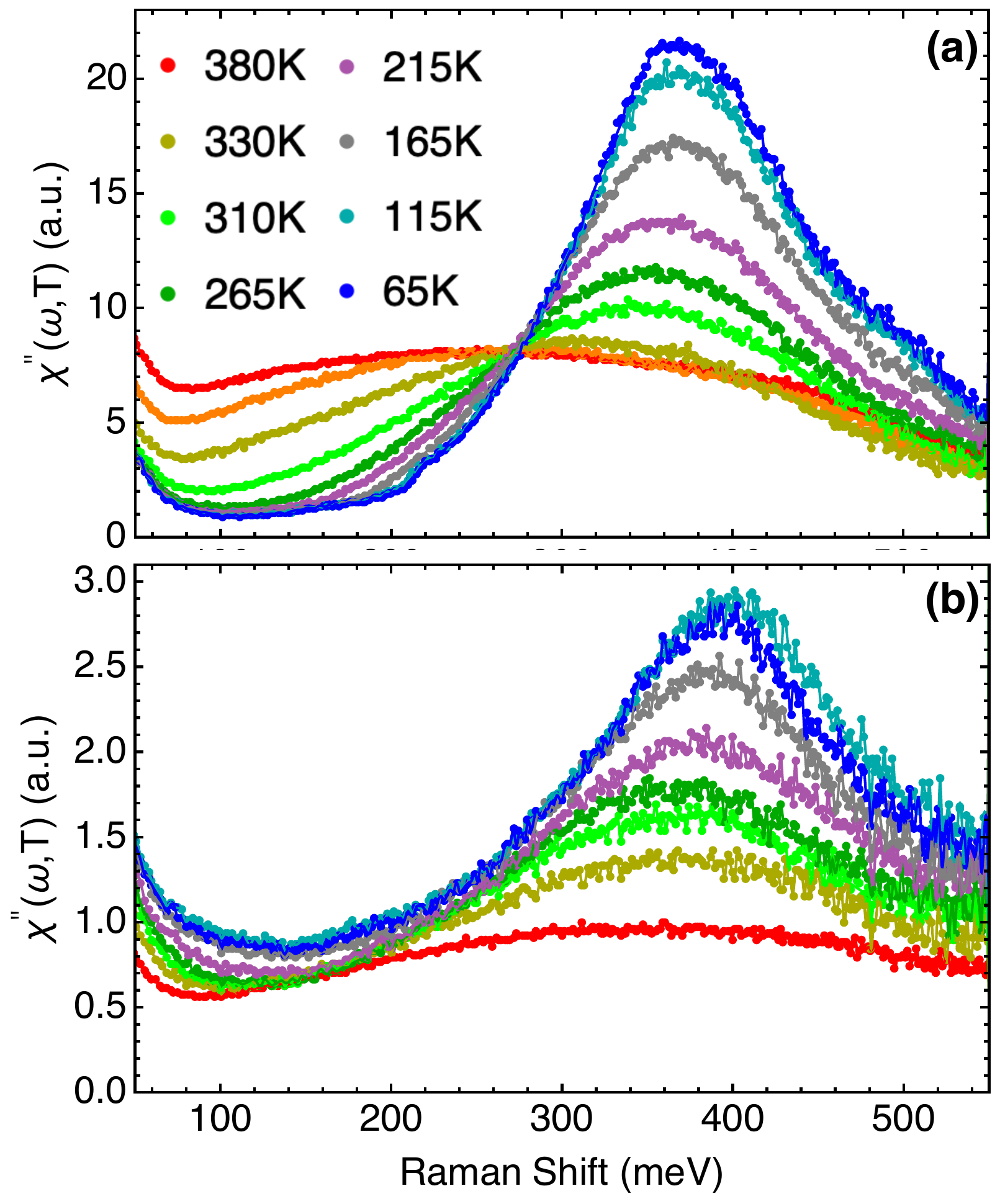}
	\caption{Temperature dependence of the high-energy Raman response in (a) $aa$ and (b) $ac$ geometries.}
	\label{fig:ac}
\end{figure}

In Fig.\,2e of the main text we compare the high-energy Raman response in $aa$ and $ac$ scattering geometries on the same scale. 
In Fig.\,\ref{fig:ac}b below we show the enlarged $ac$ spectra where the stark contrast to the $aa$ (Fig. \ref{fig:ac}a) response can be better appreciated: there is no pronounced redistribution of intensity and no gap opening on cooling. 
The intensity actually increases on cooling below $200$ meV, which can be partially attributed to "leakage" from $aa$ geometry below $T_{\mathrm{c}}$ -see subSec.\,\ref{sec:highenleak} below.

\subsubsection{Intensity redistribution in high-energy Raman spectra}

\begin{figure}[h!]
	\centering
	\includegraphics[width=0.5\textwidth]{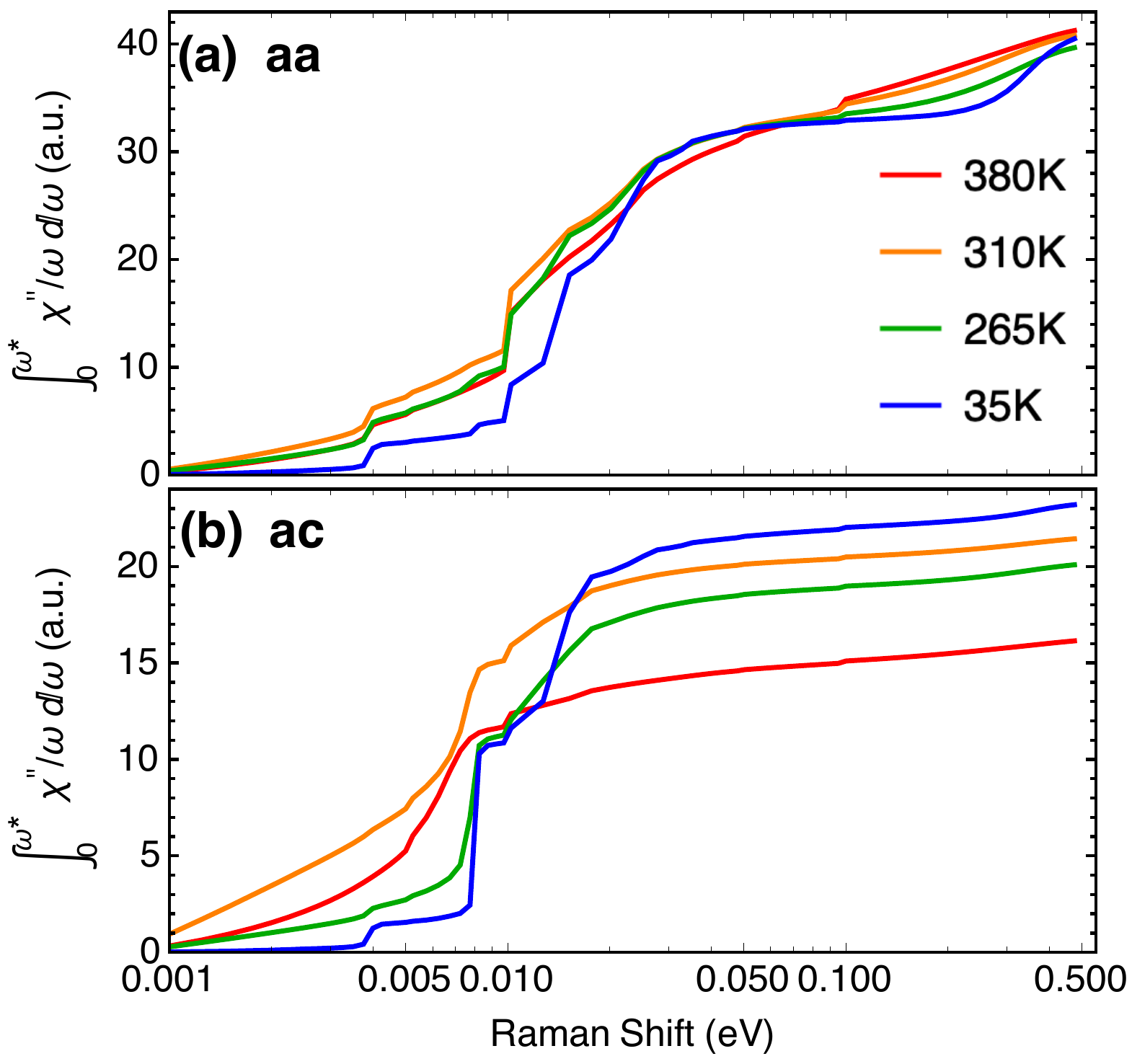}
	\caption{The integral $\int_0^{\omega*} \frac{\chi_{aa(ac)}''(\omega)}{\omega} d\omega$ as a function of the Raman shift $\omega*$ for $aa$ ({\bf a}) and $ac$ ({\bf b}) polarization geometries.}
	\label{fig:redistr}
\end{figure}

Here we illustrate the evolution of intensity in $aa$ and $ac$ geometries on cooling by computing the integral $I_{aa(ac)}(\omega^*)=\int_0^{\omega*} \frac{\chi_{aa(ac)}''(\omega)}{\omega} d\omega$ (Fig. \ref{fig:redistr}). For $\omega^*\to \infty$ this quantity is proportional to the static susceptibility in the respective geometry, while the $\omega*$ dependence allows to understand the contribution of different frequency ranges.

For $aa$ polarization (Fig. \ref{fig:redistr}a), one observes that at frequencies larger than $50$ meV, the spectral weight gets suppressed on cooling below around $300$ meV, but then recovers at $500$ meV to an almost temperature-independent value. This is consistent with the absence of strong temperature dependence of the static $A_{\mathrm{g}}$ susceptibility that has no instability in Ta$_2$NiSe$_5$. Moreover, this shows that unlike in a mean-field theory (see section S2 below) where the $A_{\mathrm{g}}$ signal should vanish in the absence of the order parameter, the integral at $500$ meV is roughly constant across $T_{\mathrm{c}}$, consistent with the presence of signatures of the excitonic pairing (the pseudogap) well above $T_{\mathrm{c}}$ observed in ARPES \cite{ARPES2014}.

In $ac$ geometry, on the other hand, $I_{\mathrm{ac}}(500 \text{meV})$ is strongly dependent on temperature. However,  $I_{\mathrm{ac}}(\omega^*)$ appear to be almost parallel above around $30$ meV, without any signs of redistribution as in $aa$ geometry. This suggests that the temperature dependence arises due to the contribution of the lower energies to the integral. Indeed, there is a strong enhancement at the lowest frequencies on cooling from $380$ K to $310$ K explained by the softening of the excitonic continuum (see Fig. 3 of the main text and Fig. \ref{fig:Linear}), and a suppression on further cooling. The enhancement on cooling from $265$ K to $35$ K can be seen to arise from energies around $15-20$ meV where a contribution of phonon "leakage" (which grows on cooling) from $aa$ geometry is prominent.

Overall, we have shown that the pronounced redistribution of the high-energy Raman response is characteristic of $aa$ geometry, but not $ac$ geometry, demonstrating that the emergence of the strong peak in the ordered phases, shown in Fig. 2e of the main text, is indeed specific for $aa$ geometry.

\subsubsection{"Leakage" in high-energy Raman spectra at low temperatures}
\label{sec:highenleak}

\begin{figure}[h!]
	\centering
	\includegraphics[width=0.5\textwidth]{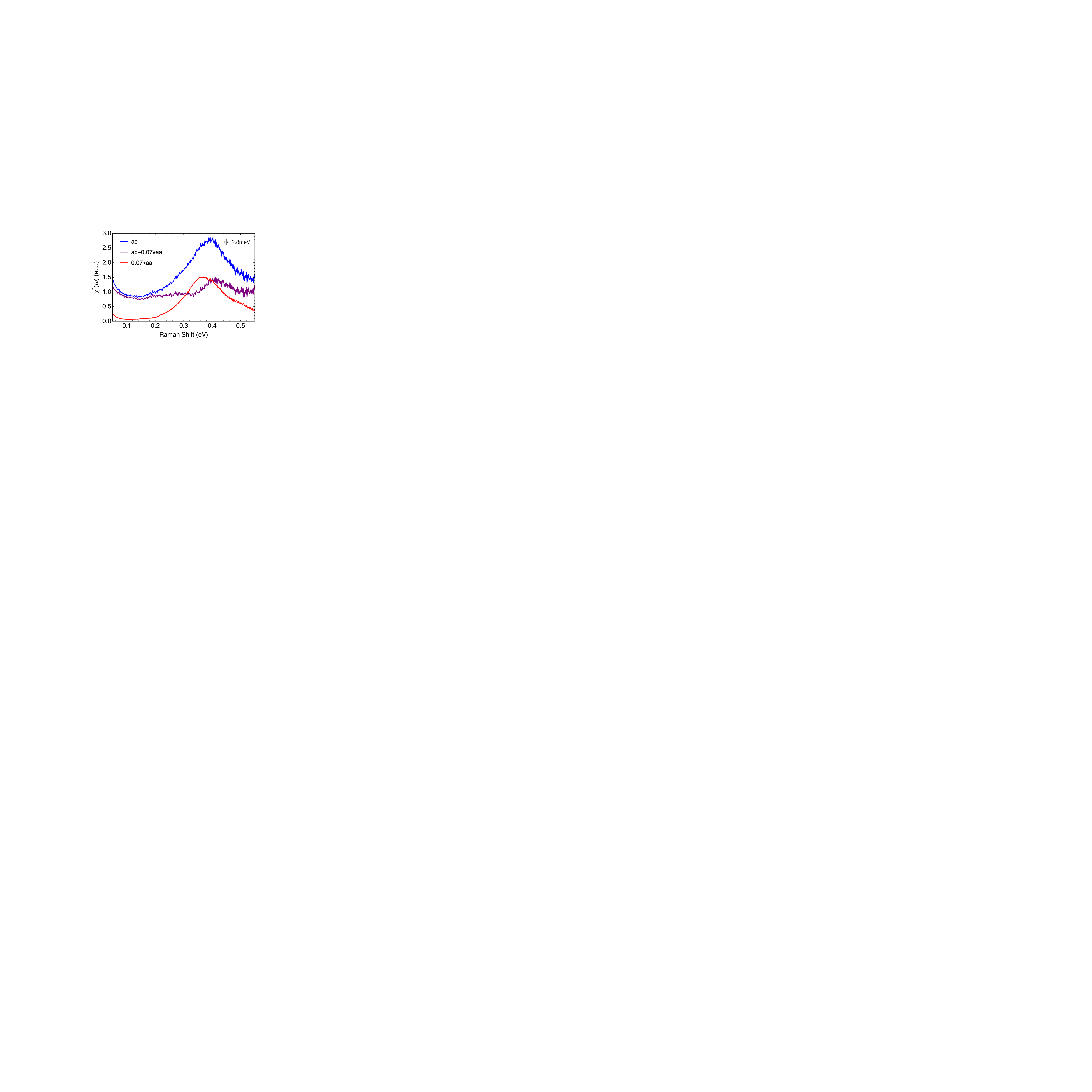}
	\caption{"Leakage" analysis of the high-energy Raman data at $35$\,K. 
		Blue line is the measured $ac$-polarization data; subtracting 7\% of the $aa$-polarization data (red line) from the $ac$-polarization data one obtains an almost flat intensity below about 0.33 eV (purple line), suggesting that there is a 7\% leakage between $aa$ and $ac$-polarized data at high-energies. The spectral resolution is 2.8\,meV.}
	\label{HE}
\end{figure}

Here we demonstrate that a large part of the $ac$ Raman intensity (Fig.\,2e) can be attributed to the change in the Raman selection rules below $T_{\mathrm{c}}$. In analogy to the phonons, one may expect the intensity from $aa$ geometry to "leak" into $ac$ geometry and vice versa. As $aa$ intensity is significantly larger in the high-energy region than the $ac$ one, we can neglect the leakage of $ac$ into $aa$, and consider only the former effect. Consequently, we have analyzed $\chi''_{\mathrm{ac}}(\omega)- p\cdot\chi''_{\mathrm{aa}}(\omega)$, where $0<p<1$. 
Intriguingly, for $p=0.07$ (Fig.\,\ref{HE}), we observe the resulting intensity (shown in purple) to be featureless below around 310\,meV, while for other values of $p$ a residual energy dependence in this region is observed.
This coincidence of lineshapes up to a background constant below around 310\,meV suggests that the "intrinsic" $ac$ intensity, corresponding to the interband transitions, is actually represented by the purple curve, Fig.\,\ref{HE}. It follows then that not only the intensity in $ac$ geometry is much weaker, but also that its lineshape is different, characterized by the appearance of intensity at higher energies than in $aa$ geometry. The latter observation is consistent with the prediction of the mean-field model, where the intensity in $aa$ geometry is expected to be peaked at the gap edge, while the one in $ac$ geometry only starts to grow at the gap edge and should be maximal at higher energies.

\section{Raman scattering in an excitonic insulator at low temperatures}
\label{app:theor}
We consider a model with an electron-like conduction band and hole-like valence band described by the hamiltonian
\begin{equation}
\hat{H}_0 = \sum_{{\bf p},\sigma} \varepsilon_{\mathrm{c}}({\bf p}) \hat{c}^\dagger_{c,\sigma}({\bf p})\hat{c}_{c,\sigma}({\bf p})+\varepsilon_{\mathrm{v}}({\bf p}) \hat{c}^\dagger_{v,\sigma}({\bf p})\hat{c}_{v,\sigma}({\bf p}),
\label{eq:H0}
\end{equation}
where the band operators at the $\Gamma$ point transform under irreducible representations $I_1$ and $I_2$ of the point group above $T_{\mathrm{c}}$ - $D_{\mathrm{2h}}$, such that $I_1\otimes I_2 \equiv B_{\mathrm{2g}}$. The latter requires the hybridization between the two bands at the $\Gamma$ point to vanish; however, below $T_{\mathrm{c}}$ the point-group symmetry is reduced to $C_{\mathrm{2h}}$ such that $B_{\mathrm{2g}}$ merges with the trivial $A_{\mathrm{g}}$ representation into a single one, allowing for the $c-v$ hybridization at the $\Gamma$ point. Thus, the quantity of the form $\langle \hat{c}_{c,\sigma}^\dagger \hat{c}_{v,\sigma'}\rangle$ may serve as the order parameter for the phase transition from  $D_{\mathrm{2h}}$ to $C_{\mathrm{2h}}$. One needs to be make sure though, that the average doesn't break any additional symmetries. In particular, no spin anisotropy or magnetism have been observed in Ta$_2$NiSe$_5$, requiring $\langle \hat{c}_{c,\sigma}^\dagger \hat{c}_{v,\sigma'}\rangle = W \delta_{\sigma,\sigma'}$. Moreover, time-reversal symmetry that acts as a complex conjugation, if spin is ignored, requires $W$ to be real. This suggests that in contrast to true bosonic condensates, there is no $U(1)$ degeneracy of the order parameter in an excitonic insulator, that is instead reduced to $Z_2$. The general reason for that is that while there is a global $U(1)$ symmetry due to particle number conservation, the number of $c$ and $v$ electrons are not separately conserved (which would yield a $U(1)\times U(1)$ symmetry otherwise), suggesting that the exciton number is not a conserved quantity, i.e.  $U(1)\times U(1)$ is initially broken to  $U(1)\times Z_2$ already in the $D_{\mathrm{2h}}$ phase. The mechanism of this breaking may be either due to pair-hopping interaction between bands, or coupling to phonons.

While most of the equations below are generic, we will use the following form of the dispersion to illustrate the results:
\begin{equation}
\varepsilon_{\mathrm{c}}({\bf p})  \approx \frac{p_x^2\pm p_0^2}{2m_{\mathrm{c}}};
\;
\varepsilon_{\mathrm{v}}({\bf p}) \approx  - \frac{p_x^2\pm p_0^2}{2m_{\mathrm{v}}},
\label{eq:disp}
\end{equation}
where with the $+$ sign dispersion is semiconducting, while it is semimetallic for $-$ sign; here $p_0$ is equal for both bands due to the charge compensation condition. While the dispersion \eqref{eq:disp} is strictly one-dimensional, we assume that the transition itself would have three-dimensional character either due to terms neglected in \eqref{eq:disp} (corresponding to interchain hopping along $b$ and $c$ directions $t_{b,c}$), or due to the coupling to optical phonons or acoustic strain fields, that are expected to have a three-dimensional dispersion. Due to the layered structure of Ta$_2$NiSe$_5$ one may also expect that the three-dimensional character would be rather weak, with an extended 2D-like regime. However, due to discrete nature of the broken $Z_2$ symmetry, this is not expected to significantly reduce $T_{\mathrm{c}}$ as even in strictly two dimensions the transition is allowed to occur.

We introduce a $2\times2$ "band space" and rewrite $\hat{H}_0$ using Pauli matrices $\tau_i$ and spinors $\Psi_{{\bf p},\sigma} = (\hat{c}_{c,\sigma}({\bf p}),\hat{c}_{v,\sigma}({\bf p}))$:
\begin{equation}
\begin{gathered}
\hat{H}_{0} = \sum_{{\bf p},\sigma} \Psi^\dagger_{{\bf p},\sigma} (E({\bf p}) +\xi({\bf p}) \tau_3)\Psi_{{\bf p},\sigma};
\\
E({\bf p}) = \frac{ \varepsilon_{\mathrm{c}}({\bf p}) +\varepsilon_{\mathrm{v}}({\bf p}) }{2};
\\
\xi({\bf p}) =  \frac{ \varepsilon_{\mathrm{c}}({\bf p}) -\varepsilon_{\mathrm{v}}({\bf p}) }{2}.
\end{gathered}
\end{equation}
In what follows we will omit $({\bf p})$ in $\xi({\bf p}) ,E({\bf p}) $, where that doesn't lead to a confusion. At low temperatures, the mean field Hamiltonian including the excitonic order parameter $W$ is
\begin{equation}
\hat{H}_{MF}=
\begin{bmatrix}
\hat{c}_{\mathrm{c}}({\bf p})
\\
\hat{c}_{\mathrm{v}}({\bf p})
\end{bmatrix}
^\dagger
\begin{bmatrix}
\varepsilon_{\mathrm{c}}({\bf p}) & W\\
W&\varepsilon_{\mathrm{v}}({\bf p})
\end{bmatrix}
\begin{bmatrix}
\hat{c}_{\mathrm{c}}({\bf p})
\\
\hat{c}_{\mathrm{v}}({\bf p})
\end{bmatrix}.
\label{eq:MFHam}
\end{equation}
Note that, unlike in a superconductor \cite{klein1984}, the mean-field hamiltonian \eqref{eq:MFHam} explicitly conserves the total particle number $N=\sum_{\bf p}\hat{c}^\dagger_{\mathrm{p}} \hat{c}_{\mathrm{p}}+ \hat{v}^\dagger_{\mathrm{p}} \hat{v}_{\mathrm{p}}$ and consequently vertex corrections for Raman scattering are not required to ensure it. Diagonalizing Eq. \eqref{eq:MFHam} , one obtains the following eigenvalues and eigenvectors:
\begin{equation}
\begin{gathered}
\varepsilon_{\pm}({\bf p}) = E({\bf p}) \pm \sqrt{\xi^2({\bf p}) +W^2};
\\
\Psi_\pm ({\bf p}) = [u_\pm({\bf p}),v_\pm({\bf p})]^T,
\\
u_\pm({\bf p}) = \pm \sqrt{\frac{1}{2}\left(1\pm\frac{\xi}{\sqrt{\xi^2+W^2}}\right)};
\\
v_\pm({\bf p}) = \sqrt{\frac{1}{2}\left(1\mp\frac{\xi}{\sqrt{\xi^2+W^2}}\right)}.
\end{gathered}
\label{eq:MFeig}
\end{equation}
The factors $u_\pm,v_\pm$ are analogous to the BCS coherence factors in a superconductor.

We consider the Raman scattering in the non-resonant approximation, where the effective coupling of the electrons to the two-photon process is described by $\sum_{\bf q} \hat{R}^{ij} {\bf A}_i({\bf q}){\bf A}_j(-{\bf q})$, where $\hat{R}^{ij} = \sum_{\bf p} \gamma^{ij}_{\alpha\beta} c^\dagger_\alpha({\bf p}) c_\beta({\bf p})$ is the Raman vertex \cite{devereaux2007}, where we additionally neglect the momentum dependence of $\gamma^{ij}_{\alpha\beta}$. The Raman susceptibility in the polarization geometry set by $i$ and $j$ is determined by the transition rate of the electronic system due to the perturbation by $\hat{R}^{ij}$. The symmetry restricts the possible form of $\hat{R}^{ij}$ \cite{devereaux2007}; for our case $\hat{R}^{aa}$ has to be of $A_{\mathrm{g}}$ symmetry and $\hat{R}^{ac}$ - $B_{\mathrm{2g}}$ symmetry. Taking these constraints into account the Raman vertices take the form
\begin{equation}
\hat{R}_{\mathrm{aa}} = \sum_p g_{\mathrm{c}} \hat{c}^\dagger_p \hat{c}_p+g_{\mathrm{v}} \hat{v}^\dagger_p \hat{v}_p
;\;
\hat{R}_{\mathrm{ac}} = g_{\mathrm{ac}} \sum_p \hat{c}^\dagger_p \hat{v}_p+  \hat{v}^\dagger_p \hat{c}_p,
\label{eq:ramvert}
\end{equation}
where $\hat{R}_{\mathrm{aa}}$ corresponds to $aa$ polarization geometry and $\hat{R}_{\mathrm{ac}}$ - to $ac$ geometry. Using Fermi's Golden rule we evaluate the probability to find:
\[
2 \pi \int |\langle+|\hat{R}|-\rangle|^2 \delta (\varepsilon_+({\bf p}) - \varepsilon_-({\bf p}) - \omega) \frac{ 2d {\bf p}}{(2\pi)^D},
\]
where we assumed that the system is fully gapped, i.e. $\varepsilon_+({\bf p}) > 0 > \varepsilon_-({\bf p})$ for all values of ${\bf p}$. Note that under this assumption the resulting expression does not depend on $E({\bf p})$.

\subsection{Semimetallic case}

In this case the bands are assumed to cross in the normal state, similar to the Fig. 1d of the main text. In this case, one can linearize the dispersion near the band crossing $\xi({\bf p}) = 0$, such that $\frac{ 2d {\bf p}}{(2\pi)^D} \approx \nu_0 d \xi$, $\nu_0$ being the density of states. Using the definition of the Raman vertex \eqref{eq:ramvert} and the eigenvectors obtained in \eqref{eq:MFeig} one gets
\begin{equation}
\begin{gathered}
I_{R_{\mathrm{ac}}} \sim  \frac{4 \pi g_{\mathrm{ac}}^2 \nu_0 \sqrt{\omega^2/4-W^2}}{\omega} \theta(\omega-2W);
\\
I_{R_{\mathrm{aa}}} \sim  \frac{\pi(g_{\mathrm{c}}-g_{\mathrm{v}})^2 \nu_0 W^2}{\omega\sqrt{\omega^2/4-W^2}} \theta(\omega-2W),
\end{gathered}
\end{equation}
where we used that $u_+v_-+u_-v_+ = \frac{\xi}{\sqrt{\xi^2+W^2}};
\;u_+u_-+v_+v_- = 0;
\;u_+u_--v_+v_-= - \frac{W}{\sqrt{\xi^2+W^2}} $ and $\delta(2\sqrt{\xi^2+W^2}-\omega) = \frac{\delta(\xi\pm\sqrt{\omega^2/4-W^2})}{2\xi/\sqrt{\xi^2+W^2}}$. Note that in the absence of the order parameter ($W=0$), the Hamiltonian \eqref{eq:MFHam} would respect the conservation of $c$ and $v$ particle numbers separately, leading to a zero Raman response in $aa$ geometry, which couples to densities of $c$ and $v$ fermions. Also, one notices that the $aa$ intensity vanishes when $g_{\mathrm{c}}=g_{\mathrm{v}}$, i.e. when the Raman vertex \eqref{eq:ramvert} becomes just the total density operator. This further shows that particle number conservation is respected in our calculation.

%{\bf We note that the transition rates are equal to twice ${\rm Im}\chi_R$, as is expected at $T=0$}.

\subsection{Semiconducting case}

In this case there is a direct gap between the two bands without hybridization. We use the 1D dispersion relation \ref{eq:disp} with the "$+$" sign, such that the direct gap is $2E_0\equiv \frac{p_0^2}{\mu}$, where $\mu=2m_{\mathrm{c}}m_{\mathrm{v}}/(m_{\mathrm{c}}+m_{\mathrm{v}})$ is the effective mass. For the Raman intensity one gets then
\begin{widetext}
\begin{gather*}
I_{R_{\mathrm{ac}}} \sim
2g_{\mathrm{ac}}^2\int \frac{(p^2/(2\mu)+E_0)^2}{(p^2/(2\mu)+E_0)^2+W^2} \delta (2\sqrt{(p^2/(2\mu)+E_0)^2/4+W^2} - \omega) dp,
\\
I_{R_{\mathrm{aa}}} \sim
\frac{(g_{\mathrm{c}}-g_{\mathrm{v}})^2}{2}\int \frac{W^2}{(p^2/(2\mu)+E_0)^2+W^2} \delta (2\sqrt{(p^2/(2\mu)+E_0)^2+W^2} - \omega) dp.
\end{gather*}
The result of the integration is:
\begin{equation}
\begin{gathered}
I_{R_{\mathrm{ac}}} \sim 2 g_{\mathrm{ac}}^2 \frac{\sqrt{2\mu}}{\omega\sqrt{\sqrt{\omega^2/4-W^2}-E_0}} \theta(\omega-2\sqrt{W^2+E_0^2}),
\\
I_{R_{\mathrm{aa}}} \sim  \frac{(g_{\mathrm{c}}-g_{\mathrm{v}})^2}{2} \frac{\sqrt{2\mu}}{\omega\sqrt{\sqrt{\omega^2/4-W^2}-E_0}} \frac{W^2}{\omega^2/4-W^2}\theta(\omega-2\sqrt{W^2+E_0^2}).
\end{gathered}
\end{equation}
\end{widetext}
Most important difference from the semimetallic case is that intensities in both geometries show singularities at the gap edge $\omega=2\sqrt{W^2+E_0^2}$. The intensity at the gap edge in $aa$ geometry is additionally multiplied by a factor $W^2/E_0^2$. Thus, for the case $E_0\ll W$ the intensity in the $aa$ geometry is expected to dominate. However, for strongly semiconducting case $E_0\gg W$, the singuarity in the $aa$ geometry is expected to be greatly suppressed by $W^2/E_0^2$, at odds to the experimental observation in Fig. 2c of the main text. While the shapes of the singularities here are related to the 1D form of the dispersion taken, we expect the qualitative arguments on the intensity dependencies on the order parameter value to hold even when the effects of non-1D dispersion are taken into account, as those are expected mostly to smear the singularity at the band edge. 

Overall, the absence of a discernible singularity in Fig. 2c in $ac$ geometry strongly favors the semimetallic state.

\subsection{Effects of a nonlocal order parameter}
The non-locality of the order in real space results in the order parameter in momentum space being momentum-dependent, i.e. $W({\bf p})$. Away from $\Gamma$ point, time reversal and inversion symmetry combined require only that $W({\bf p}) =W^*(-{\bf p})$, and thus $W({\bf p})$ can be complex, generically written as $W({\bf p}) = |W|({\bf p}) e^{i \phi({\bf p})}$, where $\phi({\bf p}) = -\phi(-{\bf p})$. Generalizing \ref{eq:MFeig} with the momentum-dependent order parameter is straightforward and results in the new eigenvalues and eigenvectors:
\begin{equation}
\begin{gathered}
\varepsilon_{\pm}({\bf p}) = E({\bf p}) \pm \sqrt{\xi^2({\bf p}) +|W|^2({\bf p})};
\\
\Psi_\pm ({\bf p}) = [u_\pm({\bf p}),v_\pm({\bf p})]^T,
\\
u_\pm({\bf p}) = \pm \sqrt{\frac{1}{2}\left(1\pm\frac{\xi}{\sqrt{\xi^2+|W|^2({\bf p})}}\right)};
\\
v_\pm({\bf p}) =  e^{-i \phi({\bf p})}\sqrt{\frac{1}{2}\left(1\mp\frac{\xi}{\sqrt{\xi^2+|W|^2({\bf p})}}\right)}.
\end{gathered}
\label{eq:MFeig2}
\end{equation}
For the semiconducting case, assuming the ${\bf p}$ dependence of $|W|({\bf p})$ does not shift the gap minimum from the $\Gamma$ point $\phi(0) = 0$ and the results for the Raman intensities are the same as described above. For the semimetallic case, however, the result changes due to the presence of a nonzero phase. Assuming the gap to open at $p_F$ and a quasi-one dimensional band structure (consistent with ARPES experiments \cite{ARPES2018TR,fukutani2019,chen2020}), we get close to the gap edge
\begin{equation}
\begin{gathered}
I_{R_{\mathrm{ac}}} \sim 4 \pi g_{\mathrm{ac}}^2 \nu_0 \left(\cos^2(\phi(p_F))\frac{\sqrt{\omega^2/4-|W|^2(p_F)}}{\omega} +\right.
\\
\left.
+\frac{1}{2} \sin^2(\phi(p_F)) \frac{\omega}{\sqrt{\omega^2/4-|W|^2(p_F)}} \right) \theta(\omega-2W);
\\
I_{R_{\mathrm{aa}}} \sim  \frac{\pi(g_{\mathrm{c}}-g_{\mathrm{v}})^2 \nu_0 |W|^2(p_F)}{\omega\sqrt{\omega^2/4-|W|^2(p_F)}} \theta(\omega-2|W|(p_F)).
\end{gathered}
\end{equation}
One observed that while $I_{R_{\mathrm{aa}}} $ has not changed, $I_{R_{\mathrm{ac}}}$ now contains a contribution, that diverges at the gap edge with its amplitude proportional to $\sin^2(\phi(p_F))$. This result is not expected to affect the results qualitatively as $p_F$ in Ta$_2$NiSe$_5$ is quite small and $\sin^2(\phi(p_F))\sim k_F^2 a_0^2$, $a_0$ being the lattice constant: for $k_F\sim 0.1$\,\AA$^{-1}$~\cite{ARPES2018TR} and $a_0=3.49$ \AA \cite{XRay2018}, one gets the estimate $k_F^2 a_0^2$ of the order $0.12\ll1$. This estimate holds for, e.g. the recently proposed \cite{millis2019} case of $W({\bf k}) = W_0 e^{ik_xa}$ (as is given in the Supplementary material to that paper). Additionally, this effect of non-locality can contribute to the apparent "leakage" of the $aa$ intensity discussed in S1.3.

\subsection{Corrections to the Raman vertex in the ordered state}

Strictly speaking, at $T<T_{\mathrm{c}}$ the irreps $A_{\mathrm{g}}$ and $B_{\mathrm{2g}}$ mix into a single $A_{\mathrm{g}}$ representation and thus the generic form of the Raman vertex operators may mix the ones at $T>T_{\mathrm{c}}$. However, this mixing is apparently rather small in the system, as the ratio of intensities in cross polarization vs. the one in parallel polarizations is below $1/10$. A possible explanation for that is that in the effective mass approximation the appearance of the order parameter does not directly affect the mass. 

%While taking other interactions into account may lead to a self energy sensitive to the order parameter a rough estimate for these corrections should be of the order $W/[{\rm bandwidth}]$. 

\subsection{Effects of disorder}

Additionally, a remnant low-energy electronic continuum contribution is observed at low temperatures in Fig. 3g of the main text. Here we show that it can be attributed to the effect of disorder. In particular, we model disorder by a varying chemical potential $V({\bf r}) \tau_0$. If considered in Thomas-Fermi approximation to \eqref{eq:MFHam}, is is seen that local $V>W$ creates a metallic "puddle"; due to the absence of translational invariance metals are expected to result in a Drude form of the Raman susceptibility at $q=0$ due to particle-hole excitations \cite{devereaux2007}, as is observed in the experiment.

\bibliography{TNS1}

\end{document}